\documentclass[12pt]{article}
\usepackage[left=2cm, right=2cm, bottom=2cm ,top = 2cm, footskip=1cm]{geometry}
\usepackage[font=small,labelfont=bf]{caption}

\usepackage{setspace}
\let\counterwithin\relax
\usepackage{chngcntr}

\usepackage{hyperref}
\usepackage{booktabs}

\usepackage[backend=biber, style=authoryear, sorting=none, natbib=true]{biblatex}
\usepackage{graphicx}
\usepackage{amsmath}

\usepackage{multirow} % for making cells that span more than 1 row in a column

\usepackage{caption} 
\usepackage{subcaption}
\usepackage{tabularx}
\newcolumntype{L}{>{\raggedright\arraybackslash}X}

\usepackage[section]{placeins}

\title{Habitual lifestyle timing explains circadian timing, but daily lifestyle changes do not, in free-living humans across 2000 days}

\author{Billy C. Smith$^{1*}$, Zeel Pansara$^{1}$, Rosanne H. Timmerman$^{1}$, Emmanuel Molefi$^{1}$,\\ 
Tiago Da Silva Costa$^{2,3,4}$, Christopher Thornton$^{5}$, Lucas G.S. França$^{6, 7}$, \\Rachel E. Stirling$^{8}$, Sarah E. Heaps$^{9}$, Philippa Karoly$^{8}$, Mario Leocadio-Miguel$^{10}$,\\ Peter N. Taylor$^{1,3,11}$, Karoline Leiberg$^{1}$, Yujiang Wang$^{1,3,11*}$}

\begin{document}
\begin{refsection}
\maketitle

% author affiliations
\begin{enumerate}
\item{CNNP Lab (www.cnnp-lab.com), School of Computing, Newcastle University, Newcastle upon Tyne, United Kingdom, NE4 5TG}
\item{Faculty of Medical Sciences, Newcastle University, Newcastle upon Tyne, United Kingdom, NE2 4HH}
\item{Northern Centre for Mood Disorders, Newcastle University, Cumbria, Northumberland, Tyne and Wear NHS Foundation Trust, Newcastle Upon Tyne, UK}
\item{National Institute for Health and Care Research, Newcastle Biomedical Research Centre, Newcastle Upon Tyne, UK}
\item{School of Computing, Engineering \& Digital Technologies, Teesside University, Middlesbrough, Tees Valley, TS1 3BX }

\item{School of Computer Science, Faculty of Science and Environment, Northumbria
University, Newcastle upon Tyne, NE1 8ST, United Kingdom}
\item{Department of Forensic and Neurodevelopmental Science, Institute of Psychiatry,
Psychology \& Neuroscience, King's College London, London, SE5 8AF, United
Kingdom}

\item{Department of Biomedical Engineering and Graeme Clark Institute, The University of Melbourne, Victoria, Australia}

\item{Department of Mathematical Sciences, Durham University, United Kingdom}

\item{School of Psychology, Faculty of Health and Wellbeing, Northumbria University,
Newcastle upon Tyne, NE1 8ST, United Kingdom}

\item{UCL Queen Square Institute of Neurology, Queen Square, London, United Kingdom, WC1N 3BG}

\end{enumerate}

\begin{center}
* Email: b.smith16@ncl.ac.uk or Yujiang.Wang@ncl.ac.uk%, Tel: 0191 208 4141 YW's number
\end{center}

\newpage
% eBioMedicine format (max 250 words)
\section*{Summary (Abstract)}

\textbf{Background}  
Both between- and within-subject variations in circadian timing matter for health. If lifestyle changes could be used to regulate circadian timing, they would offer accessible and scalable routes to chronotherapy, but this link remains unclear under real-life conditions. Here, we explore how lifestyle `traits' (such as typical wake time) and `states' (day-to-day deviations from traits, such as waking up later than typical) explain between- and within-subject variation in acrophase (peak time) of the circadian rhythm of heart rate (CRHR).

\textbf{Methods} 
We collected free-living wearable data (smartwatch, continuous glucose monitor) from healthy volunteers for up to 4 weeks. The CRHR was derived from activity-adjusted heart rate, and acrophase was defined as time-of-day at daily CRHR peak. Sleep, food, and physical activity `factors' were calculated and split into traits and states. Using a linear mixed-effects model, we tested how traits and states associate with between- and within-subject acrophase variance.

\textbf{Findings}
Data from 105 healthy volunteers (66 female, age = 42.5 $\pm$ 15.7 years) spanning \~2000 days (18.8 $\pm$ 8.30 days each) were analysed. Traits were substantially more influential than states, explaining 42.3\% versus 0.9\% of total acrophase variance. Accordingly, traits explained 86.5\% of between-subject variance, whereas states explained only 1.8\% of within-subject variance. Sleep, food and physical activity factors contributed both jointly and uniquely, and lifestyle \textit{timing} mattered most.

\textbf{Interpretation} 
Between-subject lifestyle traits explained acrophase better than within-subject lifestyle states. This asymmetry, alongside the considerable overlap between factors, supports sustained, holistic, timing-focused lifestyle adjustments as chronotherapy targets, testable through future interventional studies.

\newpage
\section{Introduction} 

The timing of the human circadian rhythm (biological clock) varies between people. Some people's rhythms run earlier than others' -- a between-subject difference generally captured by `chronotype' \citep{roenneberg_chronotype_2019}. This between-subject timing matters for health: later chronotype has been correlated with addiction, mood, sleep, and metabolic disorders \citep{knutson_associations_2018, chan_heart_2026}. An individual's circadian timing may also shift from day to day: elevated within-subject variation has been linked to increased mortality risk \citep{windred_sleep_2024} and cardiometabolic health \citep{st-onge_multidimensional_2025}; mood disorders \citep{jones_actigraphic_2005, esaki_association_2021} (accompanying symptom onset in bipolar disorder \citep{song_causal_2024}); and epilepsy \citep{smith_more_2025}.  Thus, both an individual's typical circadian rhythm timing and day-to-day stability are clinically relevant, yet both are hard to measure at scale outside of laboratory conditions. However, the heart has its own circadian rhythm (CRHR), regulated by the SCN \citep{hastings_clockwork_2003, lu_mammalian_2021}, which can be measured with wearable devices \citep{ bowman_method_2021, huang_distinct_2021, natarajan_circadian_2025,}. The `acrophase' of this rhythm, the time of its daily peak \citep{esaki_association_2021, natarajan_circadian_2025}, can serve as a scalable proxy for circadian timing. Identifying what factors affect CRHR acrophase is a key step towards regulating between- and within-subject circadian timing in targeted chronotherapy.

Circadian rhythm timing is driven by `zeitgebers' (time-givers) \citep{aschoff_exogenous_1960} via entrainment. The light-and-dark cycle is the primary zeitgeber, but elements of lifestyle can act as non-photic zeitgebers \citep{mistlberger_nonphotic_2005, healy_circadian_2021}. Such lifestyle zeitgebers make for attractive intervention targets, as they are measurable and modifiable at scale: delayed (made later) sleep shifts circadian timing and increases timing variability \citep{sletten_timing_2010}; meal timing entrains circadian rhythms independently of sleep, with delayed meals delaying timing \citep{wehrens_meal_2017, krauchi_alteration_2002, quante_zeitgebers_2019, yoshizaki_influence_2013}; and morning physical activity generally advances (make earlier) timing, whereas evening activity delays it \citep{youngstedt_human_2019, thomas_circadian_2020, quante_zeitgebers_2019}. However, these findings come almost entirely from controlled laboratory studies, where one cue is manipulated and the rest held constant. If chronotherapeutic interventions are to take place outside the clinic, we need to understand how lifestyle factors jointly influence circadian timing under real-world conditions.

A further unknown is whether lifestyle acts mainly on between- or within-subject circadian variation, or both. Answering this means separating between- from within-subject effects in one model. The within-between framework \citep{bell_explaining_2015, mundlak_pooling_1978} does so by splitting each lifestyle predictor into an individual mean (trait), capturing habitual lifestyle, and daily deviations from that mean (state), capturing day-to-day change. This lets us ask two questions of every factor: (i) does a person's habitual lifestyle relate to their typical acrophase, and (ii) do daily lifestyle deviations relate to daily shifts in acrophase? Either answer would be informative: a factor may govern typical acrophase, its daily variation, both, or neither.

Here, we ask how much lifestyle habits, and daily deviations from them, explain between- and within-subject variance in the acrophase of the circadian rhythm of heart rate (CRHR), with multiple lifestyle factor categories modelled concurrently. We first test whether a large healthy cohort shows substantial between- and within-subject variance in CRHR acrophase using wearable devices. We then ask how much variation is accounted for by each category (sleep, physical activity, and food), and whether habit and daily deviation contribute differently between and within people. By showing which factors associate with between-subject variation, and which explain day-to-day within-subject changes, these analyses indicate which lifestyle factors are the most plausible targets for chronotherapy. \newline

%----------------------------------------------------
% METHODS
%----------------------------------------------------

\newpage
\section{Methods} % see Suppl_methods.tex for old version

\begin{figure}[h!]
    \centering
    \includegraphics[scale=1]{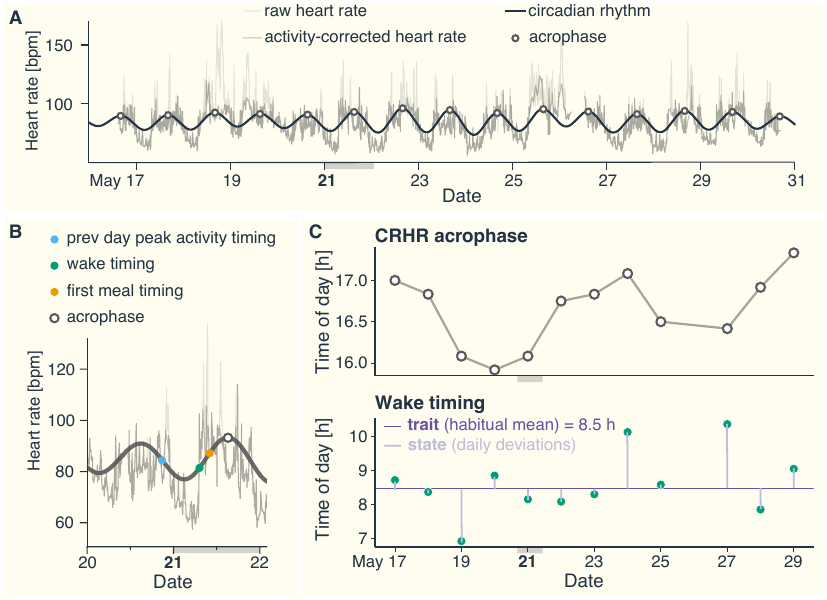}
    \caption{\textbf{Overview of Methods}
    \textbf{A}: Raw and activity-corrected heart rate over 14 days, with singular spectrum analysis (SSA) extracted circadian rhythm (circadian rhythm of heart rate, CRHR) overlaid. Empty circles indicate daily CRHR acrophase. \textbf{B}: Two-day zoom-in showing acrophase on one day together with a sample of lifestyle factors across categories (previous day peak activity time, wake time, first meal time). \textbf{C}: Top: CRHR acrophase shown as time of day over days, which is our response variable for LMER modelling. Bottom: Wake timing as an example lifestyle factor. Each factor is also split into its trait (habitual mean) and state (daily deviation) predictor factor for LMER modelling. Shaded x-axis region shows the same date across all three panels (21st May). }
    \label{fig:methods}
\end{figure}

\subsection{Study design and participants}
We collected wearable data from healthy adult volunteers in an observational, free-living design, in which participants were asked to maintain their usual routines while continuously wearing a smartwatch and a continuous glucose monitor (CGM) for up to four weeks. Glucose time-in-range for participants with CGM is shown in Supplementary~\ref{section:gluc-tir}. Recruitment was from the local area (Newcastle upon Tyne, UK); full details, inclusion criteria and device models are given in Supplementary Methods (Supplementary~\ref{section:supp_methods}). Our analysis sample comprised 105 participants contributing 1977 circadian cycles. The study was approved by the Newcastle University Ethics Committee (23-032-WAN).

\subsection{Estimating circadian timing from heart rate}
The body's circadian timing is governed not by a single clock but by a hierarchy: a central clock in the suprachiasmatic nucleus (most directly indexed by dim-light melatonin onset), and peripheral clocks throughout the body's tissues that are also entrained directly by non-photic cues such as feeding and physical activity. Our marker of interest is the circadian rhythm of heart rate (CRHR), extracted from continuous heart-rate recordings, whose daily peak time (\textit{acrophase}) we take as a readout of peripheral clock timing (Figure~\ref{fig:methods}A). CRHR acrophase is related to, but distinct from, both central clock phase and `chronotype' as captured by questionnaires \citep{vitale_chronotype_2015, roenneberg_chronotype_2019}. CRHR phase has shown reasonable alignment with DLMO, the gold standard central clock marker \citep{huang_distinct_2021}. We interpret CRHR acrophase as a wearable-derived marker of peripheral circadian timing, informative in its own right \citep{bowman_method_2021, huang_distinct_2021, lee_real-world_2024} with clinical relevance \citep{jones_actigraphic_2005, esaki_association_2021, song_causal_2024, smith_more_2025}. 

Because physical activity raises heart rate independently of the circadian rhythm (masking), heart rate during periods of detected physical activity were corrected before rhythm extraction. The circadian rhythm was extracted using singular spectrum analysis (SSA), a non-parametric method validated for extracting a circadian rhythm from physiological time-series data against alternative techniques \citep{molefi_chronossa_2026}. Acrophase was computed for each day, allowing it to vary from one day to the next (Figure~\ref{fig:methods}). Its relation to sunrise and sunset is shown in Supplementary~\ref{section:sunset-sunrise}. Full preprocessing, extraction parameters, and circadian-component identification are detailed in Supplementary Methods.

\subsection{Lifestyle factor categories}
We measured several lifestyle factors (Figure~\ref{fig:methods}B) across three categories. 

\textbf{1 Sleep}: Sleep timing, duration, and quality were derived as factors from device sleep-detection output. 

\textbf{2 Physical activity}: Bouts of physical activity events were derived from heart rate (HR), step count, and accelerometry. We included the total duration and average HR (a marker of intensity) of bouts within two hours of waking, or two hours of sleep on the previous day. Additionally, we included the timing, duration, and average HR of the bout with the greatest average HR (`peak-activity') on the previous day. 

\textbf{3 Food}: Meal timing factors were derived by glucose profile peaks as a proxy. We additionally measured pre-bed and post-wake glucose concentration as additional physiological factors. 

Detection algorithms, thresholds, and the full factor list are given in Supplementary Methods; for readability, factors are referred to by descriptive name in the text, with machine-readable names tabulated in Supplementary~\ref{section:supp_main_coefficients}.

\subsection{Separating habit (trait) from daily deviation (state)}
Our central analytical step separates each (lifestyle) factor into two components, following the within-between framework \citep{bell_explaining_2015, mundlak_pooling_1978}. The \textbf{trait} component reflects a person's average level of a lifestyle factor across their recording, capturing habits (e.g typical wake time). The \textbf{state} component is their daily deviation from that average, capturing day-to-day change (e.g waking up 1 hour later than usual); also see Figure~\ref{fig:methods}C. Modelling both together lets us ask two distinct questions of every factor: does a person's \textit{habit} relate to when their circadian rhythm typically peaks (a between-subject question), and do \textit{daily deviations from habits} relate to daily shifts in rhythm timing (a within-subject question)? This separation is what dense wearable sampling makes possible, and where it differs from questionnaire instruments, which capture habitual between-subject difference, but not its within-subject day-to-day movement. See Supplementary~\ref{section:supp_definitions} for a full list of paper terminology used.

\subsection{Statistical analysis}
We modelled the effects of lifestyle factors (predictors) on CRHR acrophase (response) using a linear mixed-effects regression with fixed effects for all lifestyle (trait and state), and a random intercept for participant. Other factors were also included as predictors including participant age, sex, and daylight hours for any given day. Two smartwatches were used in our study: to correct for device-level confounds, device type was included as a factor.  We confirm that circular (time of day, 0h to 23.99h) acrophase and lifestyle factors are linearly associated in Supplementary~\ref{section:supp_local_lin}. A first-order autoregressive structure was placed on the residuals to absorb day-to-day circadian inertia, so that the variance of estimates of the lifestyle coefficients can be directly quantified. To describe how much acrophase variance lifestyle explained, we used marginal and conditional $R^2$ for overall fixed- and random-effect fit \citep{nakagawa_general_2013}, and semi-partial $R^2$ \citep{stoffel_partr2_2021} to attribute variance to each lifestyle category uniquely and to the variance shared across categories. Finally, we used a Snijders-Bosker decomposition \citep{snijders_multilevel_2011} to split the variance explained into its between- and within-subject parts. 
To guide interpretation, each analysis was mapped to a specific question: marginal and conditional $R^2$ quantified the overall variance explained by the model; semi-partial $R^2$ quantified the unique and shared contributions of lifestyle categories; grouped semi-partial $R^2$ compared the explanatory contribution of trait versus state factors; the Snijders-Bosker decomposition separated explained variance into between- and within-subject components; and standardised $\beta$ coefficients were used to compare the relative direction and magnitude of individual lifestyle factors. All model specifications, variance decomposition formulae, and software are given in Supplementary Methods.

%----------------------------------------------------
% RESULTS
%----------------------------------------------------
\newpage
\section{Results}
We analysed wearable data from 105 healthy, free-living, volunteers (66 female, age = 42.5 $\pm$ 15.7 years) over 1977 circadian cycles (days) (18.8 $\pm$ 8.30 cycles each) (Supplementary~\ref{section:supp_demographics}). 

\subsection{Between- and within-subject heart rate circadian acrophase distributions}

\begin{figure}[h]
    \centering
    \includegraphics[scale=1]{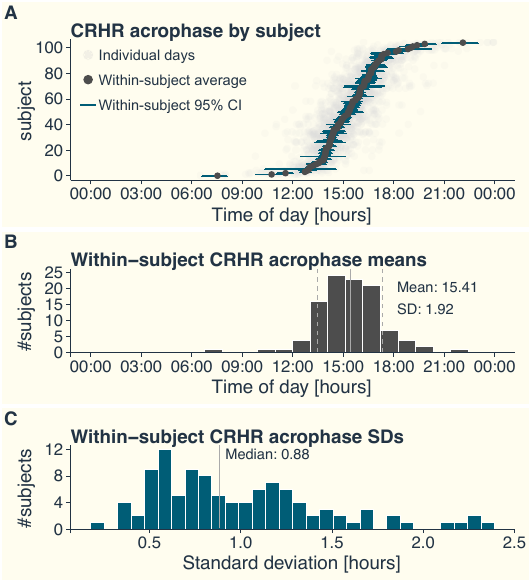}
    \caption{\textbf{CRHR acrophase distributions within and between subjects.}
    \textbf{A}: Caterpillar plot of per-subject CRHR acrophase distributions.
    \textbf{B}: Distribution of within-subject acrophase average (circular mean). \textbf{C}: Distribution of within-subject acrophase standard deviations.}
    \label{fig:acro_dist}
\end{figure}

% Finding
We first investigated the distribution of acrophase for each subject (Figure \ref{fig:acro_dist}A), and extracted within each subject their mean and standard deviation (SD) in CRHR acrophase (Figure \ref{fig:acro_dist}B, C). 
% Magnitude
Across all subjects, their mean acrophase was distributed on average at 15.41~hours (03:25pm) with a standard deviation of 1.92~hours (Figure \ref{fig:acro_dist}B). 
Across all subjects, their acrophase standard deviations (Figure \ref{fig:acro_dist}C) range from 0.21 hours to 2.35 hours (median 0.88~h).
% Biological Interpretation
Between-subject variability in average acrophase (1.92h) was greater than the median within-subject variability in acrophase (0.88h) in this healthy volunteer population. Having established that acrophase varies substantially both between and within individuals, we now ask how much of each source of variation is explained by lifestyle.

\subsection{60.65\% of CRHR acrophase variance can be explained}
\begin{figure}[h]
    \centering
    \includegraphics[scale=1]{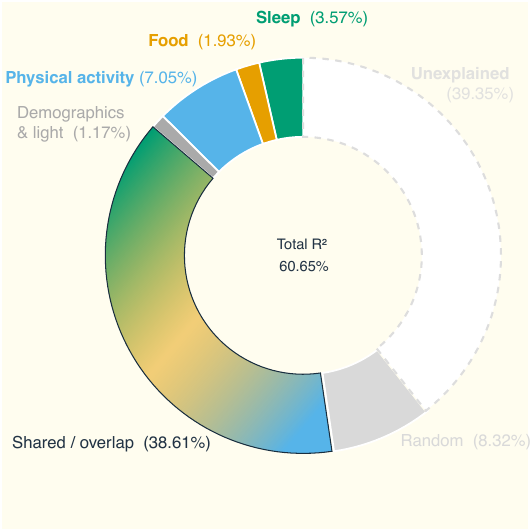}
    \caption{% Declarative Title
    \textbf{60.65\% of CRHR acrophase variance can be explained by modelled factors.}
    Coloured portions represent variance explained (semi-partial $R^2$) by categories of lifestyle factors included in the model. Distinct lifestyle categories sleep, food and physical activity uniquely explained 3.57\%, 1.93\%, 7.05\%, respectively. Modelled demographics and light contributed a further 1.17\%. The overlap between lifestyle categories explained 38.61\% of variance. The random intercept further explains 8.32\% variance, leaving a total of 39.35\% variance unexplained. 
    }
    \label{fig:maindonut}
\end{figure}

% Finding
We investigated the extent to which lifestyle factor categories (sleep, food, physical activity) explain CRHR acrophase variance overall. Using semi-partial $R^2$, we evaluated both the unique contributions of each category, the overlapping contributions from all lifestyle categories, non-lifestyle categories (including factors of demographics and light), and the random intercept (Figure~\ref{fig:maindonut}).
While each category also uniquely contributed to explain the CRHR acrophase variance, we found that the overlap between categories -- representative of variance not uniquely attributable to any group, due to inherent associations between categories -- explained the largest proportion of variance attributable to the lifestyle model.

% Magnitude (include marginal, conditional $R^2$)
The model total marginal (fixed effects only) and conditional (fixed and random effects) $R^2$ were 52.33\% and 60.65\% respectively. 
Of the investigated model components, the shared variance explained between categories (`overlap') contributed most (38.61\%) to the model marginal $R^2$. Individual categories uniquely contributed between 1.93\% (food) and 7.05\% (physical activity). The model's random intercept, capturing between-subject variability not covered by traits, was associated with 8.32\% variance. 39.35\% acrophase variance remains unexplained by our model. 

% Biological Interpretation
The large portion of variance explained attributed to the overlap implies that these lifestyle categories are not strictly independent zeitgebers, but form a correlated lifestyle profile. Despite their inherent overlap, there are small, but clear portions of acrophase variance uniquely attributable to different lifestyle categories.

\subsection{Traits explain CRHR acrophase better than states}

\begin{figure}[h!]
    \centering
    \includegraphics[scale=1]{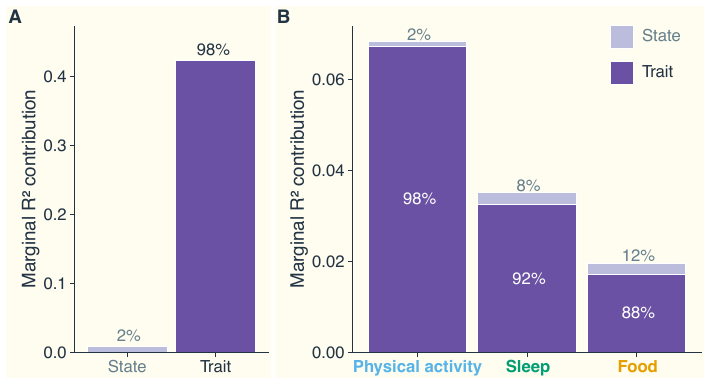}
    \caption{  % Declarative Title
    \textbf{Lifestyle traits explained substantially more variance than states. }
    % Essential Methods 
    Semi-partial $R^2$ (the reduction in model marginal $R^2$ when a specific group of factors are dropped) was used to determine the marginal $R^2$ contributions shown. Percentages show the relative contributions between traits and states.
    %  Statistical Information
     \textbf{A}: Variance explained by trait and state factors uniquely. \textbf{B}: Same as A, but restricted to unique variance explained within each lifestyle category.
    }
    
    \label{fig:traitstate}
\end{figure}

% Finding
To contextualise the previous result, we investigated the differences in variance explained by lifestyle trait (individual habitual mean) and state (deviation from habit) separately. We found that across and within all lifestyle categories, traits were substantially more influential than states in explaining CRHR acrophase variance (Fig.~\ref{fig:traitstate}). Together, all lifestyle trait factors uniquely accounted for $R^2=0.423$ variance explained. In comparison, state factors only accounted for $R^2=0.009$ (Fig.~\ref{fig:traitstate}A). This is a split of approx. 98\% \textit{vs.} 2\% of relative explainable variance contributions.
Within each lifestyle category (Fig.~\ref{fig:traitstate}B), traits accounted for between 88-98\% of relative variance. Conversely, states accounted for between 2-12\%. This implies that modelled habits (such as typical wake time) are more strongly associated with CRHR acrophase than modelled day-to-day deviations from habits (e.g., later/earlier wake times relative to typical).
Furthermore, we performed a Snijders-Bosker \citep{snijders_multilevel_2011} multi-level decomposition (Section \ref{section:snijders-bosker}) to separate variance in CRHR acrophase into within- \textit{vs.} between-subject aspects. We found that while 86.5\% ($R^2_{between} = 0.865$) of between-subject CRHR acrophase variance was explained by the model, only 1.8\% ($R^2_{within}=0.018$) of within-subject CRHR acrophase was explained. Taken together, these findings imply that modelled between-subject habitual lifestyle traits explain between-subject CRHR acrophase variance well, but within-subject lifestyle states mostly do not explain within-subject day-to-day CRHR acrophase variance. Put simply, using wake-time as an example: a person's habitual wake time explains their average CRHR acrophase well, but daily variations in wake time do not explain their daily changes in acrophase.

\subsection{Timing-related traits have the greatest influence on CRHR acrophase}

\begin{figure}[h!]
    \centering
    \includegraphics[scale=1]{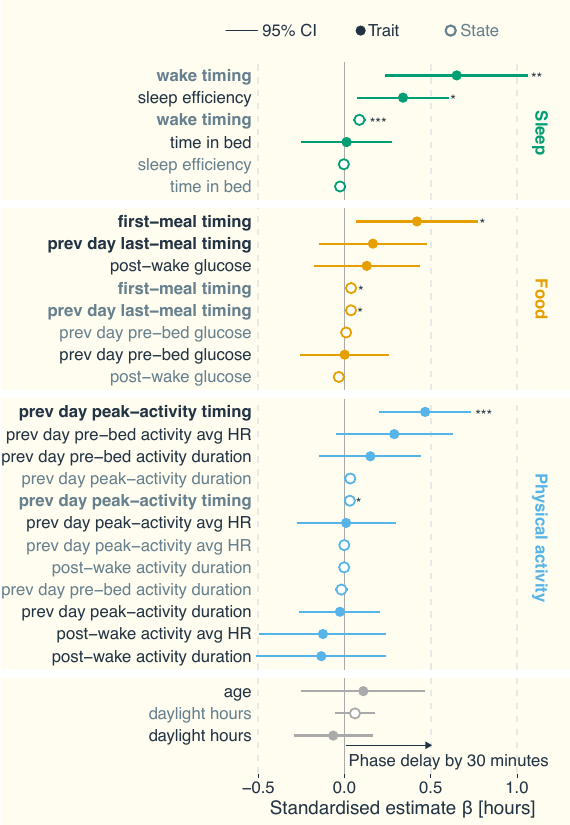}
    \caption{
    \textbf{Magnitude of influence for individual factors.} Points show standardised model estimates ($\beta$, beta weights), with 95\% confidence intervals overlaid as bars. For example, $\beta=0.5$ means that a factor, when \textit{increased} by one unit of standard deviation, delayed (made later) acrophase by 0.5h = 30 minutes. Negative $\beta$ indicate acrophase advance (make earlier) for each one standard deviation \textit{increase}. Timing-related factors are highlighted in bold text. Traits are shown with a filled circle, and states are shown with an empty circle. Approximate two-sided p-values (* = $<$ 0.05) are shown for reference only. 
    }
    \label{fig:indi_estimates}
\end{figure}

% Finding 
Finally, we evaluated the importance of individual lifestyle factors on CRHR acrophase through comparison of model coefficients (Figure~\ref{fig:indi_estimates}), which reflect how much acrophase is advanced or delayed per one standard deviation increase in each factor, when other factors are held constant. There was considerable variation between factors in the estimate magnitude and direction, from -0.13 hours (0.13h advance) for post-wake activity duration (trait), to 0.65 hours (0.65h delay) for wake timing (trait). Standard error also varied from 0.01 hours for previous day's peak-activity average heart rate (state) to 0.21 hours for wake timing (trait). Timing-related trait factors had the greatest influence on CRHR acrophase. 

% Magnitude
Overall, individual factors had a weak influence on CRHR acrophase, most within the range of $\beta$ = 0.0~h to 0.5~h (advance or delay to acrophase) for each one standard deviation increase in the factor. The most influential factors fell between 0.5~h to 1.0~h, with none exceeding 1.0~h. Note the standard deviation of each factor is its population standard deviation across subjects and days. Across categories, timing-related trait factors had the greatest association with CRHR acrophase, specifically the factors of `wake time', `first meal time', and `previous day's peak-activity time'. %Some state factors showed considerable influence however, such as `wake time', and `post-wake activity duration and average heart rate'. 
A detailed table of model coefficients is shown in Supplementary~\ref{section:supp_main_coefficients}. 

The wide confidence intervals around several estimates indicate uncertainty in the precise size, and in some cases direction, of these associations. We investigated this further via a leave-one-subject-out sensitivity analysis. While there was substantial variation in the estimate of most coefficients, most maintained their general direction (Supplementary~\ref{section:supp_LOSO}).

To summarise, the most influential factors were timing-related habitual (trait) factors. The model’s explanatory power is not driven by one dominant lifestyle factor or category, but from multiple related timing cues that together define an individual’s circadian timing.

\subsection{Modelled lifestyle categories account for weekday-weekend differences in CRHR acrophase}

% Finding
In our model, there was not a substantial weekday \textit{vs.} weekend effect on acrophase ($\beta= 0.002 \pm 0.04$, see Supplementary~\ref{section:supp_main_coefficients}). We investigated this in further detail in Supplementary~\ref{section:supp_weekend}. Most subjects did not have a substantial difference in acrophase between weekdays and weekends, though the number of subjects with a $\geq 30 min$ later acrophase on weekends was greater than the number with a $\geq 30 min$ earlier weekend acrophase. When re-running the model over weekdays-only and weekends-only, trends seen in Figures~\ref{fig:maindonut} and \ref{fig:indi_estimates} generally held. However, sleep and overlap explained relatively more variance during weekends, and conversely, physical activity and the random intercept explained more variance during weekdays.

\section{Discussion}

% Para 1 -- What it means (the state-trait asymmetry as central finding)
The timing of an individual's circadian rhythm has two distinct properties: when it typically peaks, which differs markedly between people, and how much timing shifts from day to day. Using \~2000 days of continuous wearable recordings from 105 healthy volunteers under free-living conditions, we found that everyday lifestyle accounts for these two properties very differently. Habitual lifestyle explained most of the between-subject differences in CRHR acrophase: people whose habits are generally early or late have circadian rhythms that are accordingly early or late. Yet the same lifestyle factors explained almost none of a person's day-to-day changes in acrophase, despite daily sampling that could resolve any day-to-day effects precisely. Beyond this asymmetry, two further features stand out: lifestyle factors act far more through their shared, overlapping influence than through any single zeitgeber acting alone; and it is the \textit{timing} of lifestyle, rather than quality or quantity, that was associated with circadian timing. Together they reframe how lifestyle should be understood as a regulator of our biological clock. \newline

% Para 2 -- Evidence and prior work 
The magnitudes we observed are consistent with circadian shifts reported under controlled conditions, but our free-living design reframes their interpretation. A one-standard-deviation delay in habitual sleep-wake time delayed acrophase by 0.65 $\pm$ 0.21 hours, comparable to the 1.75-hour delay \cite{sletten_timing_2010} reported for a two-hour sleep delay under laboratory control. Habitual first meal timing showed the same direction as the 1.1--3.8 hour delays of \cite{yoshizaki_influence_2013}, though weaker (0.42 $\pm$ 0.18). This is unsurprising: \cite{yoshizaki_influence_2013} delayed all three daily meals equally, and controlled caloric intake (a considerable aspect of meals as a zeitgeber \citep{stephan_calories_1997}), whereas we observed meals as they naturally occurred. Glucose peak timing may also be a less reliable proxy. Physical activity followed the expected geometry: habitually more intense post-wake activity advanced phase (-0.12 $\pm$ 0.19 hours per standard deviation), more intense pre-bed activity delayed it (0.29 $\pm$ 0.17 hours), and later timing of the day's most intense bout delayed acrophase (0.47 $\pm$ 0.14 hours), echoing the general morning-advance, evening-delay pattern of \cite{thomas_circadian_2020} and \cite{quante_zeitgebers_2019}. In summary, we find that effects established one cue at a time in the laboratory survive when every cue varies at once, but in attenuated form. 

Under free-living conditions, \cite{quante_zeitgebers_2019} report stronger shifts from later meal and physical activity than we observed, however report independent correlations where lifestyle factors were not mutually adjusted. In contrast, we were able to describe the unique contributions of each lifestyle factor or category when considered alongside other factors/categories (Figure \ref{fig:maindonut}). Lifestyle factors measured in real-world settings have inherent interrelationships (Supplementary~\ref{section:supp_vifs}), which we were able to cleanly separate from unique contributions through describing the variance explained by `overlap'. Crucially, no single factor or category dominated; lifestyle factors are not competing zeitgebers, but a co-acting profile: they shift circadian timing together. This explains the large overlap and motivates treating lifestyle holistically rather than as a set of independent levers, a point we return to below.    \newline 

% Para 3 -- State-trait asymmetry interpretation, with causality caveat
What, then, does the state-trait asymmetry imply about the CRHR? That habitual lifestyle explains when the rhythm typically peaks, while day-to-day deviation barely tracks how it changes day-to-day, suggests it is shaped by sustained lifestyle patterns rather than by short-term fluctuation. Stable patterns, such as a wake time held at 6am rather than 8am over weeks, appear to matter; sporadic late wake times do not. This parallels evidence that the central clock remains stable over months despite lifestyle variation \citep{mchill_robust_2021}, and raises the possibility that peripheral timing is similarly buffered against daily perturbation. It also helps reconcile our attenuated effects with the larger shifts reported in the laboratory: a controlled study imposes a large, sustained, single-cue shift, whereas natural day-to-day variation is small and potentially self-cancelling. So controlled laboratory results are not overstating an effect, but measuring a response to a much larger and cleaner perturbation. The direction of this relationship under real-world conditions, however, remains open. Our observational design cannot separate lifestyle acting as a zeitgeber from the endogenous central clock that drives both peripheral physiological phase and the timing of lifestyle; the within-between model lets us hold habitual timing and daily deviation apart, but it describes association, not cause. \newline

% Para 4 -- Clinical implications
In terms of clinical implications, our results point to concrete targets to test for causal chronotherapy interventions. Because habitual, timing-related factors carried the most weight, the most promising levers for shifting peripheral rhythm timing are sustained adjustments to the timing of sleep, meals, and exercise rather than other properties or any occasional change. To advance (shift earlier) a person's circadian rhythm, a regimen of progressively earlier wake, meal, and activity times maintained over weeks, even if light conditions are not controlled, is the approach our findings would support. The dominance of shared over unique variance carries a practical implication: because lifestyle factors act in concert, chronotherapeutic adjustment is likely to be more effective when applied holistically. However, this combined causality also needs to be tested in interventional studies. Factors not tied to voluntary timing may offer secondary targets: pre-bed routines to improve sleep efficiency, dietary structure to regulate glucose, or activity prescribed to target specific heart-rate zones. An additional open question is whether there is an interaction between trait and state: for example whether deviations from typical wake time are more disruptive to circadian timing for someone who wakes early, compared to someone who wakes late. We were unable to widely test interaction terms in our model due to sample size limitations, but preliminary testing did not reveal any substantial interaction between a trait and its respective state (Supplementary~\ref{section:supp_interactions}), suggesting this may not be the case.  All of this, however, follows from association, and the causal step must be demonstrated directly through interventional studies before any of it is clinically actionable \citep{quante_zeitgebers_2019}. \newline

% Para 5 -- Open questions / future work (reframed from limitations)
Several questions follow directly from what we found, and from what our design could not reach. The first is direction: establishing whether habitual lifestyle entrains the peripheral circadian rhythm, or merely co-varies with an endogenous clock that sets both, requires interventional designs that manipulate habitual timing and observe the rhythm's response. The second is the large unexplained share of within-subject variance: day-to-day acrophase shifts substantially, yet the lifestyle factors we measured account for almost none of that movement, even though we were well powered to detect it across \~2000 person-days. This potentially points to inputs we did not capture (principally light exposure, but also mood \citep{shapiro_unraveling_2024}, stress \citep{steinach_circadian_2020}, alcohol \citep{meyrel_alterations_2020}, caffeine \citep{burke_effects_2015}), social factors \citep{mistlberger_social_2004}, and work schedules \citep{bowman_method_2021}. Emerging light-tracking and ecological momentary assessment methods could close some of this gap. Additionally, it may point towards between-subject differences in how lifestyle states interact with acrophase (for example, meal timing's effect on acrophase being influenced by between-subject differences in metabolism or food intake \citep{stephan_calories_1997}).  A third concerns generalisability of the marker and the sample. CRHR acrophase reflects one peripheral circadian rhythm and need not track rhythms in other domains (core temperature, melatonin, or cellular clocks). Our heart-rate correction mitigates the primary masking confound of physical activity but not of somnolence, sleep debt, or pathology. Our cohort, though spanning a broad age range, was predominantly white and drawn from one region. Finally, the between-subject findings rest on \~100 individuals; while our daily sampling resolves within-subject effects well, the habitual estimates and the between-person variation in their direction would be sharpened by larger cohorts. \newline

% Para 6 -- Conclusion
In conclusion, across \~2000 days and 105 healthy adults, lifestyle explained when the circadian rhythm of heart rate typically peaks far better than how it shifts from day to day. We detected a strong shared influence of many factors over any single cue. The timing of lifestyle appeared to matter more than any other property. The asymmetry we report between habitual and day-to-day timing is the step that lets the next questions be posed precisely: which lifestyle timing cues, adjusted deliberately and sustained over weeks, actually shift the rhythm in a causal study design? What governs the day-to-day variations in circadian rhythm that habitual lifestyle leaves unexplained? Answering them would turn these associations into the basis for targeted chronotherapy, and clarify why circadian timing becomes more variable in neurological and psychiatric disease.

\section*{Software}
R (4.6.0) \citep{r_core_team_r_2025}, packages: Rssa (1.0.4) \citep{korobeynikov_rssa_2024}, nlme (3.1-169) \citep{pinheiro_nlme_2026}, MuMIn (1.48.19) \citep{barton_mumin_2026}, performance (0.17.0) \citep{ludecke_performance_2021}

\section*{Code and Data availability}

Code and data to reproduce our full statistical analysis is provided in \url{https://github.com/cnnp-lab/2026_BCS_LifestyleCRHR}.

\section*{Acknowledgements}
We thank all study volunteers, and all members of the Computational Neurology, Neuroscience \& Psychiatry Lab (\url{www.cnnp-lab.com}) for discussions on the analysis and manuscript.

P.N.T. and Y.W. are both supported by UKRI Future Leaders Fellowships (MR/T04294X/1, MR/V026569/1).

\section*{Author contributions}

B.S, K.L, Y.W. contributed to study design, interpretation, and writing. 

B.S, R.T, E.M, K.L, Y.W. contributed to data analysis. 

Z.P, B.S, C.T and Y.W. contributed to data collection. 

B.S, Y.W, P.N.T. contributed to figure design. 

Y.W, K.L, P.N.T, P.K, M.L, T.D.S.C, S.H, R.S, Z.P, E.M, L.G.F, C.T, and R.T. contributed to paper review. 

\section*{Conflict of interest}
None declared.

\newpage
\printbibliography[title={References}]
\end{refsection}

%%%%%%%%%% %%%%%%%%%% SUPPLEMENTARY %%%%%%%%%% %%%%%%%%%% 

\newpage

%%%%%%%%%% %%%%%%%%%% SUPPLEMENTARY %%%%%%%%%% %%%%%%%%%% 

\newpage
\renewcommand{\thefigure}{S\arabic{figure}}
\captionsetup[table]{name=Supplementary Table}
\captionsetup[figure]{name=Supplementary Figure}
\setcounter{figure}{0}
\setcounter{equation}{0}
\counterwithin{figure}{section}
\counterwithin{equation}{section}
% \counterwithin{table}{section}
% \renewcommand\thesection{S\arabic{section}}
\setcounter{section}{0}

\begin{refsection}
\section*{Supplementary Sections}

\section{Glucose time-in-range}\label{section:gluc-tir}

%Purpose
Participants wore a Freestyle Libre 2 or 3 plus continuous glucose monitor (CGM) to monitor interstitial glucose levels. No participant declared a diabetes diagnosis to the study team. The quality of an individual's glucose control can be approximated by duration of time interstitial glucose readings are between 3.9mmol/L (70 mg/DL) and 10mmol/L (180mg/DL) (referred to as `time in range' (TIR) \citep{bellido_time--range_2021}). \cite{beck_validation_2018} report an average TIR from blood measurements across 1440 people with type 1 diabetes of 41\% ($\pm$ 16\%). Most people with type 1 or type 2 diabetes using CGMs are recommended to aim for a TIR above 70\%. In Supplementary Figure~\ref{figure:supp_gluc-tir}, we evaluated the TIR of our participants to ensure we were sampling from a non-diabetic population. 
%Finding
With a minimum of 90\%, the time-in-range of study participants is typical of previous observations for non-diabetic populations \citep{shah_continuous_2019}; substantially higher than what has been observed in diabetic populations \citep{beck_validation_2018}; and well above the target range of 70\% for people with diabetes \citep{bellido_time--range_2021}.
%Conclusion
%This indicates our study population are most likely either non-diabetic, or their diabetes is well-controlled. 

\begin{figure}[h!]
    \centering
    \includegraphics[scale=1]{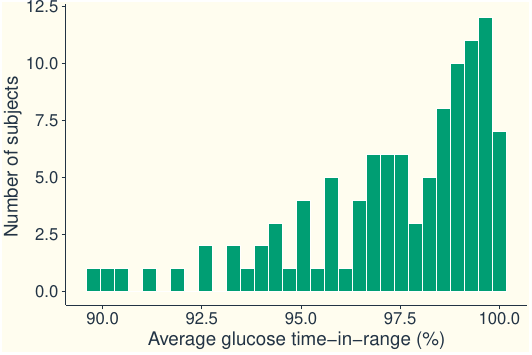}
    \caption{The percentage of recording time where participants' interstitial glucose levels (measured using CGM) were in the range of 3.9mmol/L to 10mmol/L (70-180 mg/DL). }\label{figure:supp_gluc-tir}
\end{figure}

\newpage
\section{Supplementary Methods}\label{section:supp_methods}

\subsection{Participant data}
Wearable device data were collected from healthy volunteers in an observational design. Adult participants were recruited through various means (colleagues, mailing lists, posters, online advertisements, word-of-mouth) from the local area (Newcastle upon Tyne, UK). Collection and analysis of this data was approved by the Newcastle University Ethics Committee (23-032-WAN). Study inclusion criteria at signup were age $>$ 18, generally of good health, willing to wear wearable devices and be prompted with digital questionnaires for up to four weeks, not aware of health issues preventing continuous use of wearable devices (including overnight wear). Participants were asked to maintain their usual routines during the four week experimental period. Participants wore a smart watch (Empatica EmbracePlus or Garmin Venu 3) and optionally a continuous glucose monitor (`CGM'; Freestyle Libre 2 or 3 plus; glucose time-in-range shown in Supplementary~\ref{section:gluc-tir}). Smartwatches collected heart rate (via photoplethysmography, PPG), various measures of accelerometry-derived physical activity, and sleep characteristics derived from proprietary device algorithms. Garmin data were collected via the `Labfront' platform.  
For inclusion in this analysis, additional pre-processing criteria were CGM data present, age and sex recorded, no reporting of shift work, and no travel outside of a UTC+2 timezone. Individual cycles were dropped where smartwatch data coverage was less than 75\% of the day, or acrophase was outlying ($>$5h from subject median); subjects where all cycles were dropped where removed.  Overall, 105 participants (73 with a Garmin, 32 with an Empatica, all with a CGM) were retained for analysis here.

\subsection{Lifestyle feature extraction}

\subsubsection{Sleep detection}

Sleep windows were derived separately for each device and unified from properietary algorithm outputs into a common format. 
For Garmin: sleep onset, wake, and duration were taken directly from the Labfront-exported daily summary. 
For Empatica, sleep was detected from the device's sleep-detection stage output (stage codes: 0 = awake, 101 = sleep, 102 = brief awakening, 300 = prolonged awakening).  Contiguous (with maximum allowed gap of 30 minutes) non-awake blocks were clustered to form candidate sleep episodes; the longest candidate per night was selected, and the time of sleep onset and wake derived from it.
`Time in bed' (TIB) refers to the duration between onset and wake. Duration was defined as the total time actually spent sleeping (TIB, with awakenings removed). Efficiency was calculated as $ duration\over{TIB}$.

\subsubsection{Physical activity detection}

An algorithm was developed to derive `Physical activity' (PA) events from raw heart rate (HR), step count, and accelerometry data recorded by both Garmin and Empatica devices at 1-minute sample rate. Firstly, missing intervals in wearable data shorter than 5 minutes were linearly interpolated and smoothed using a symmetric triangular kernel (11-minute window for HR for higher temporal precision; 21-minute window for step count and accelerometry, for greater smoothing). Activity types were identified via thresholds: walking (50–120 steps/min), running ($\geq$120 steps/min), `other training' (HR $>$50\% of age-predicted maximum ($(220 - age) \times 0.5$), and acceleration $>$ 2 $\times$ individual mean - possibly reflective of weight training, or non-walking aerobic exercise such as swimming or cycling). When overlapping, walking and running events took precedence over `other training'. Events under 5 minutes were excluded.

\subsubsection{Glucose profile peak detection}

Glucose profile peak times, interpreted as a proxy for meal times, were identified from the ~15-minute CGM time series using a template matching approach \citep{pellizzari_automatic_2025}. A 2-hour Gaussian template ($\sigma$ = 20 minutes) was cross-correlated with a 24-hour rolling-baseline-corrected glucose signal. Candidate peaks were required to exceed a minimum cross-correlation score of 5 and prominence of 10 mmol/L-normalised units, with a minimum inter-peak separation of 2 hours. For each peak, the post-peak trough was identified within a 3-hour forward search window, and peak height (peak glucose - preceding trough) was calculated.

\subsection{Circadian rhythm of heart rate and acrophase estimation}

\subsubsection{Preprocessing}

For each subject, heart rate (HR) data were resampled to 5-minute resolution (mean of 5-minute bins) from an original 1-minute resolution common to both devices, to reduce noise and computational time. Small gaps ($\leq$ 1 hour) were filled by linear interpolation. HR data were then split into variable-length `runs' between missing-data gaps exceeding 60 hours (2.5 days), and each run was centred on the within-run mean by subtraction. Runs were required to have at least 2.5 days of gap-free data. Very few subjects had more than 1 run, and no subjects had more than 3 runs. 
To minimize the effect of PA-induced HR elevations when extracting a circadian rhythm, HR data bins covering detected PA events were removed from the run. 
Any remaining missing gaps ($>$ 1 hour, or associated with PA) in the run were interpolated using an iterative gap-filling approach based on singular spectrum analysis (SSA) (igapfill; \citep{korobeynikov_rssa_2024}). 

%
%To correct HR for physical activity (PA), HR signal corresponding to PA bins (with $\pm$10 minutes padding) were treated as missing and reconstructed using iterative SSA-based gap-filling prior to final decomposition, so that the effect of PA-induced HR elevations were minimized in the extracted rhythm.

\subsubsection{Circadian rhythm extraction}

After resampling and interpolation, singular spectrum analysis (SSA) (Rssa \citep{korobeynikov_rssa_2024}) was applied to each HR run to extract a circadian rhythm. 
SSA is a non-parametric spectral estimation technique that can be used to separate underlying distinct-frequency rhythmic components from noise in time-series data. It is an additive decomposition, such that all the distinct-frequency `components' extracted by SSA can be added together to produce the input data. For example, if a set of distinct-frequency sine waves and a noise component were added together to produce a mixed-frequency, noisy signal, passing this through SSA should return the original set of sine waves, and an identified noise component. In this way, it is useful to extract underlying biological rhythms (including, but not limited to, the circadian rhythm) from real-world, noisy biological time-series data. Minimal parameters are required, only a `window length' parameter ($L$) which can be considered as an `upper bound' on the frequency of any extracted components. 
An additional advantage of SSA over traditional circadian rhythm extraction techniques -- such as cosinor analysis -- is that it better captures the day-to-day fluctuations in circadian rhythm properties \citep{molefi_chronossa_2026} such as acrophase.
For SSA, $L$ was set to 720 5-minute samples (2.5 days). Edge artefacts were mitigated by forecast-padding both ends of the data before application of SSA, then dropping the sections of the output corresponding to padding. 
To identify the component corresponding to a circadian rhythm from the set of distinct-frequency components, we first ordered components by the proportion of overall signal variation explained (eigenvalue). For each component, we used the fast fourier transform (FFT) to estimate the spectral power (the sum of squared absolute FFT coefficients in a given frequency range) in: the component overall and the circadian range (20 to 28 hour period). If most ($\geq$ 60\%) of the components total power fell under the 20 to 28 hour range, the component was classified as circadian. The circadian time series (Figure \ref{fig:methods}A) for each run was reconstructed as the sum of identified circadian components (if multiple were detected), plus the run mean.

\subsubsection{Acrophase}

For each subject, a singular daily acrophase time series was constructed (Figure \ref{fig:methods}). Across the circadian time series computed for each run, acrophase was calculated for each day of recording as the time-of-day (in decimal hours, e.g 3:30pm = 15.5) in local time at circadian rhythm peak magnitude. %Acrophase was not computed for days in-between extracted runs. 

\subsection{Statistical analysis}\label{stats}

\subsubsection{Modelled lifestyle factors}

Lifestyle factors were modelled at daily resolution and fell into one of three broad categories of lifestyle factors: sleep, food, or physical activity. Factor names are prefixed by category in the following statistical model description.
`Non-behavioural' (`other') factors were also modelled, including demographic factors (age and sex), a boolean weekend factor (is\_weekend, i.e did that cycle fall on a saturday or sunday), corrections for smart watch device type (deviceGarmin or deviceEmpatica), and the amount of available daylight on that day in Newcastle upon Tyne based on the duration between sunrise and sunset (capturing seasonal variation).
A full list of the 45 factors included in the linear mixed effects model is shown in Supplementary~\ref{section:supp_main_coefficients}.

\subsubsection{Modelling habits and day-to-day deviation: trait and state decomposition}

Following the `within-between' framework proposed by \cite{bell_explaining_2015} (building upon \cite{mundlak_pooling_1978}), each lifestyle factor was decomposed at the individual-level to the mean across their observation period, reflecting subject-specific lifestyle habits, and daily deviations from their mean, reflecting within-subject day-to-day deviation from lifestyle habits. 
Including both components simultaneously in a mixed-effects model allows the between- and within-subject lifestyle-acrophase effects to be estimated simultaneously; with the additional benefit that unobserved between-subject confounds are absorbed by the between-subject means, so are removed from the day-to-day deviations. 
Borrowing from psychology terminology \citep{schmitt_statetrait_2020}, we term these decompositions `trait' and `state' factors respectively. Figure \ref{fig:methods}C shows variation in lifestyle factor `wake time' for one individual over 12 days, which is decomposed into trait and state when modelled. In this instance, the wake time trait -- the average wake time for this individual -- is 8:28am. The state is a series of within-subject daily deviations from this average, calculated by subtracting the trait from the original factor, which captures earlier or later wake times each day. The trait would be modelled as a constant over the 12 observations.

\subsubsection{Binary missingness indicators}\label{section:binary_missing}

When a lifestyle factor was missing on a given day, the corresponding state factor was set to 0 and an associated missingness indicator factor was set to 1. This prevents missing data from being interpreted as days on which behaviour was exactly average — a state value of 0 ordinarily means no deviation from the person's trait. By including the missingness indicator as a covariate, the model estimates a distinct effect for missing days rather than conflating them with observed days where the factor aligned with the average.

For example, an individual may have forgotten to wear their watch on one night, meaning sleep factors (such as wake time) were not recorded. However, their CGM was still attached, and they put their watch back on after waking, before morning exercise. Rather than drop this entire day from the model, losing valid morning meal time and activity data, we set the sleep state factors to 0 and sleep\_data\_missing to 1.

\subsubsection{LMER model}

Modelling the influence of behaviour on CRHR acrophase was performed using a linear mixed-effects regression (LMER) model with fixed effects for lifestyle and non-lifestyle factors (see Supplementary~\ref{section:supp_main_coefficients} for a full list of included factors), and a random intercept for individual.  
We used the R NLME implementation of an LMER (nlme::lme \citep{pinheiro_nlme_2026}) which additionally allows modelling of a first-order autoregressive correlation structure (corAR1) on model residuals, indexed by calendar day within participant. This absorbs day-to-day `circadian inertia' into the residual covariance structure rather than conditioning on yesterday’s acrophase as a fixed effect, so that the variance of estimates of the lifestyle coefficients can be properly quantified. Due to the extensive number of factors, a summary of the formula for the model is shown below:
\begin{equation}
    \begin{split}
   lme\_full: circadian\_acrophase \sim & sleep\_factors \\
    & + food\_factors \\
    & + activity\_factors\\
    & + other\_factors\\
    & + corAR1\\
    & + (1|subject) 
    \end{split}
    \label{eqn:monthly_MELR_reduced}
\end{equation}
A key assumption of linear mixed effects models is that the predictors are linearly related to the response factor. However both circadian acrophase, as well as timing predictor factors, are circular (time-of-day, 0h to 23.99h). Supplementary~\ref{section:supp_local_lin} shows the correlation between these factors, confirming that they are `locally' linearly related, so suitable for LMER modelling.
The final sample for LMER modelling included 1977 circadian cycles (days) across 105 participants.

\subsubsection{Variance decomposition}\label{section:variance-decomposition}

To summarise the variance in acrophase explained by our model, we used marginal and conditional $R^2$. Marginal $R^2$ ($mR^2$) refers to variance explained by LMER fixed effects only, whereas conditional $R^2$ ($cR^2$) refers to the variance explained by fixed and random effects together \citep{nakagawa_general_2013}. We used the MuMIn R package \citep{barton_mumin_2026} to calculate marginal and conditional $R^2$.

To compare the acrophase variance explained by lifestyle categories uniquely (i.e, within distinct groups of the fixed effects), we calculated semi-partial $R^2$ \citep{stoffel_partr2_2021}, calculated by leave-one-block-out refitting. For example, to calculate the semi-partial $R^2$ ($spR^2$) of only factors in the sleep category, we calculate:

\[spR^2_{sleep} = mR^2 - mR^2_{no\_sleep}\]

where $mR^2$ refers to the marginal $R^2$ of the main model with all lifestyle categories, and $mR^2_{no\_sleep}$ refers to the marginal $R^2$ of a sub-model with all sleep factors removed. If sleep is important for explaining acrophase variance, then  $mR^2_{no\_sleep}$ should be lower than $mR^2$. The magnitude of the reduction in variance explained is the semi-partial $R^2$ for the sleep category, and specifically represents the variance uniquely explained by that group of factors.

If each group of fixed effects are perfectly independent, the sum of their semi-partial $R^2$s (unique variance explained) should sum to the total variance explained by all fixed effects in the model together (marignal $R^2$). But what if the sum is smaller than the marginal $R^2$? It is possible that some variance cannot be uniquely attributed to any group, but is `shared' between two or multiple groups. i.e, when one group of factors is dropped to calculate its $spR^2$, another group `steps in' to account for some of the variance that dropped group explained. In this case, there is `overlap' in the way that different groups of lifestyle factors explain acrophase variance. Note that semi-partial $R^2$s still account for the unique variance explained by each group, but their sum will be less than the model $mR^2$, as some variance is not uniquely attributable to any group. To measure this, we calculate: 
 
\[spR^2_{overlap} = mR^2 - (spR^2_{sleep} + spR^2_{food}  + spR^2_{pa} + spR^2_{other})\]

Finally, the variance associated with the model random intercept uniquely is calculated as $spR^2_{random} = cR^2 - mR^2$, and the model unexplained variance is calculated as $spR^2_{unexplained} = 1 - cR^2$.

\subsection{Snijders-Bosker multilevel analysis}\label{section:snijders-bosker}
We used the Snijders-Bosker \citep{snijders_multilevel_2011} multi-level decomposition to estimate the between- and within-subject variance explained by lifestyle factors. To do so, we compared the main model ($lme\_full$, Eqn.~\ref{eqn:monthly_MELR_reduced}) to a `null' model ($lme\_null$, Eqn.~\ref{eqn:monthly_MELR_null}), without any lifestyle traits or states:

\begin{equation}
    \begin{split}
   lme\_null: circadian\_acrophase \sim & other\_factors\\
    & + corAR1\\
    & + (1|subject) 
    \end{split}
    \label{eqn:monthly_MELR_null}
\end{equation}

For both models, we extracted the intercept ($\tau^2$, between-subject) and residual ($\sigma^2$, within-subject) variance components from the model fits. We then calculated the variance explained ($R^2$) at either the between-subject or within subject level by lifestyle factors (traits and states):

\[R^2_{between} = 1 - \frac{\tau^2_{lme\_full}}{\tau^2_{lme\_null}} \]

\[R^2_{within} = 1 - \frac{\sigma^2_{lme\_full}}{\sigma^2_{lme\_null}}\]

Compared to a null model with no lifestyle factors, this result indicates that lifestyle factors are able to account for 86.5\% of all between-subject variance in acrophase. Conversely, lifestyle factors are only able to account for 1.8\% of within-subject variance.

This implies that the within-subject variance in acrophase remains unexplained by the included factors. Together with our finding in Figure~\ref{fig:traitstate} that the trait factors are substantially more influential than state factors, these results imply that between-subject acrophase variation can be explained by lifestyle traits, but within-subject variance cannot be explained by lifestyle states.

\newpage
\section{Acrophase phase relationship with sunset and sunrise}\label{section:sunset-sunrise}
\begin{figure}[h!]
    \centering
    \includegraphics[scale=1]{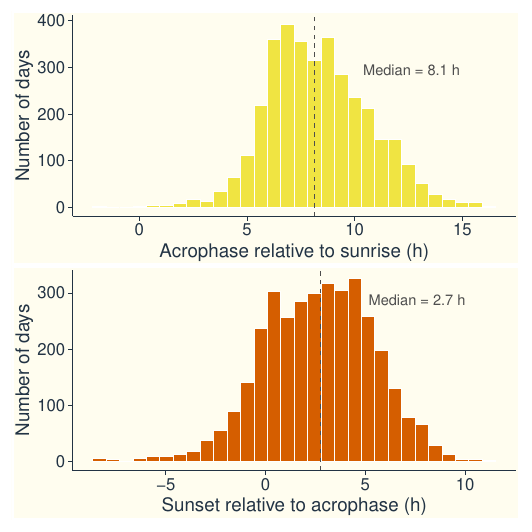}
    \caption{The relationship between CRHR acrophase and sunrise/sunset timing (in Newcastle upon Tyne) across subjects.}\label{fig:supp_sunrise_sunset}
\end{figure}

To additionally motivate CRHR acrophase as a circadian phase marker, we provide distributions of the phase relationship in Figure \ref{fig:supp_sunrise_sunset}. Acrophase occurred approximately 8.1 hours after sunrise. Acrophase occurred approximately 2.7 hours prior to sunset.

\newpage
\section{Full list of model coefficients}\label{section:supp_main_coefficients}

  \begin{table}[h!]
        \centering

        \scalebox{0.75}{
           \begin{tabular}{l|l|l|r|r|r|r|r}
                \hline
                category & effect & term & estimate & std.error & df & statistic & p.value\\
                \hline
                n/a & fixed & (Intercept) & 14.9134065 & 0.2856241 & 1845 & 52.2134122 & 0.0000000\\
                \hline
                other & fixed & age & 0.1091742 & 0.1834586 & 86 & 0.5950892 & 0.5533465\\
                \hline
                other & fixed & sexMale & 0.3173780 & 0.2201307 & 86 & 1.4417704 & 0.1529989\\
                \hline
                other & fixed & is\_weekendTrue & 0.0021677 & 0.0373377 & 1845 & 0.0580575 & 0.9537091\\
                \hline
                other & fixed & deviceGarmin & 0.4299571 & 0.3459302 & 86 & 1.2429014 & 0.2172822\\
                \hline
                other & fixed & daylight\_hours\_trait & -0.0653622 & 0.1175184 & 86 & -0.5561870 & 0.5795273\\
                \hline
                other & fixed & daylight\_hours\_state & 0.0607888 & 0.0596867 & 1845 & 1.0184642 & 0.3085909\\
                \hline
                sleep & fixed & sleep\_waketime\_trait & 0.6504174 & 0.2115630 & 86 & 3.0743438 & 0.0028271\\
                \hline
                sleep & fixed & sleep\_waketime\_state & 0.0865741 & 0.0180912 & 1845 & 4.7854265 & 0.0000018\\
                \hline
                sleep & fixed & sleep\_tib\_trait & 0.0121599 & 0.1336144 & 86 & 0.0910071 & 0.9276986\\
                \hline
                sleep & fixed & sleep\_tib\_state & -0.0255195 & 0.0159729 & 1845 & -1.5976730 & 0.1102870\\
                \hline
                sleep & fixed & sleep\_efficiency\_trait & 0.3390633 & 0.1359634 & 86 & 2.4937830 & 0.0145531\\
                \hline
                sleep & fixed & sleep\_efficiency\_state & -0.0037453 & 0.0148075 & 1845 & -0.2529339 & 0.8003475\\
                \hline
                sleep & fixed & sleep\_data\_missing1 & -0.4517538 & 0.3211113 & 1845 & -1.4068450 & 0.1596418\\
                \hline
                food & fixed & gluc\_postwake\_trait & 0.1287704 & 0.1565334 & 86 & 0.8226383 & 0.4129889\\
                \hline
                food & fixed & gluc\_postwake\_state & -0.0330013 & 0.0184991 & 1845 & -1.7839455 & 0.0745968\\
                \hline
                food & fixed & gluc\_daybefore\_prebed\_trait & 0.0008196 & 0.1314049 & 86 & 0.0062369 & 0.9950381\\
                \hline
                food & fixed & gluc\_daybefore\_prebed\_state & 0.0094409 & 0.0152645 & 1845 & 0.6184881 & 0.5363300\\
                \hline
                food & fixed & gluc\_firstmeal\_time\_trait & 0.4202630 & 0.1809266 & 86 & 2.3228373 & 0.0225493\\
                \hline
                food & fixed & gluc\_firstmeal\_time\_state & 0.0381228 & 0.0164522 & 1845 & 2.3171763 & 0.0206031\\
                \hline
                food & fixed & gluc\_daybefore\_lastmeal\_time\_trait & 0.1647770 & 0.1593647 & 86 & 1.0339617 & 0.3040529\\
                \hline
                food & fixed & gluc\_daybefore\_lastmeal\_time\_state & 0.0377388 & 0.0147918 & 1845 & 2.5513274 & 0.0108114\\
                \hline
                food & fixed & gluc\_postwake\_missing1 & -0.0947042 & 0.0883154 & 1845 & -1.0723405 & 0.2837074\\
                \hline
                food & fixed & gluc\_daybefore\_prebed\_missing1 & 0.0601319 & 0.0890563 & 1845 & 0.6752121 & 0.4996256\\
                \hline
                food & fixed & gluc\_firstmeal\_missing1 & -0.0615598 & 0.0851569 & 1845 & -0.7228987 & 0.4698337\\
                \hline
                food & fixed & gluc\_firstmeal\_nopeak1 & 0.0110882 & 0.0469861 & 1845 & 0.2359893 & 0.8134672\\
                \hline
                food & fixed & gluc\_daybefore\_lastmeal\_missing1 & 0.0334467 & 0.0872534 & 1845 & 0.3833285 & 0.7015203\\
                \hline
                food & fixed & gluc\_daybefore\_lastmeal\_nopeak1 & -0.0335989 & 0.0455204 & 1845 & -0.7381061 & 0.4605439\\
                \hline
                pa & fixed & pa\_postwake\_duration\_trait & -0.1344254 & 0.1918486 & 86 & -0.7006847 & 0.4853913\\
                \hline
                pa & fixed & pa\_postwake\_duration\_state & -0.0019437 & 0.0151898 & 1845 & -0.1279609 & 0.8981939\\
                \hline
                pa & fixed & pa\_postwake\_avghr\_trait & -0.1249903 & 0.1878474 & 86 & -0.6653821 & 0.5075867\\
                \hline
                pa & fixed & pa\_daybefore\_prebed\_duration\_trait & 0.1499550 & 0.1508396 & 86 & 0.9941354 & 0.3229459\\
                \hline
                pa & fixed & pa\_daybefore\_prebed\_duration\_state & -0.0177773 & 0.0180266 & 1845 & -0.9861702 & 0.3241789\\
                \hline
                pa & fixed & pa\_daybefore\_prebed\_avghr\_trait & 0.2883500 & 0.1723710 & 86 & 1.6728446 & 0.0979921\\
                \hline
                pa & fixed & pa\_daybefore\_maxbout\_time\_trait & 0.4673701 & 0.1358087 & 86 & 3.4413842 & 0.0008956\\
                \hline
                pa & fixed & pa\_daybefore\_maxbout\_time\_state & 0.0310887 & 0.0138502 & 1845 & 2.2446424 & 0.0249092\\
                \hline
                pa & fixed & pa\_daybefore\_maxbout\_avghr\_trait & 0.0093738 & 0.1463884 & 86 & 0.0640336 & 0.9490920\\
                \hline
                pa & fixed & pa\_daybefore\_maxbout\_avghr\_state & -0.0012074 & 0.0130528 & 1845 & -0.0925019 & 0.9263093\\
                \hline
                pa & fixed & pa\_daybefore\_maxbout\_duration\_trait & -0.0266930 & 0.1197493 & 86 & -0.2229073 & 0.8241363\\
                \hline
                pa & fixed & pa\_daybefore\_maxbout\_duration\_state & 0.0334082 & 0.0173797 & 1845 & 1.9222600 & 0.0547269\\
                \hline
                pa & fixed & pa\_postwake\_missing1 & -0.0285459 & 0.0522429 & 1845 & -0.5464080 & 0.5848516\\
                \hline
                pa & fixed & pa\_postwake\_nobout1 & 0.0372550 & 0.0383238 & 1845 & 0.9721116 & 0.3311225\\
                \hline
                pa & fixed & pa\_daybefore\_prebed\_missing1 & 0.0333984 & 0.0618864 & 1845 & 0.5396736 & 0.5894873\\
                \hline
                pa & fixed & pa\_daybefore\_prebed\_nobout1 & -0.0243621 & 0.0389563 & 1845 & -0.6253715 & 0.5318046\\
                \hline
                pa & fixed & pa\_daybefore\_maxbout\_missing1 & 0.0165777 & 0.0590799 & 1845 & 0.2805982 & 0.7790500\\
                \hline
                pa & fixed & pa\_daybefore\_maxbout\_nobout1 & 0.0884162 & 0.0706666 & 1845 & 1.2511737 & 0.2110297\\
                \hline
            \end{tabular}
    }
    \caption{Coefficients for the intercept and 45 factors included in the model. Coefficient estimate and standard errors are in units of hours. }\label{tab:main_coefficients}
    
    \end{table}

\newpage
\section{Definition of terms}\label{section:supp_definitions}

\begin{table}[h!]
    \centering
    \begin{tabularx}{\linewidth}{L|L}
         \textbf{Term} & \textbf{Definition} \\ \hline 
         \textit{HR} & heart rate  \\ \hline
         \textit{CRHR} & circadian rhythm of heart rate \\ \hline
         \textit{PA} & physical activity\\ \hline
         \textit{Acrophase} & time-of-day at CRHR peak on a given day; a day-to-day marker of circadian phase \\\hline
         \textit{Lifestyle} &  quantifiable aspects of daily living, such as sleep, food and physical activity characteristics.\\\hline
         Lifestyle \textit{factor} & quantifiable behavioural or non-behavioural aspect of lifestyle, such as sleep timing, glucose levels post-wake, or physical activity duration.  \\\hline
         Lifestyle factor \textit{category} & a group of lifestyle factors, such as multiple factors related to sleep. We use three categories in this paper: \textit{sleep, food, physical activity}\\\hline
         Lifestyle \textit{trait} & within-subject average of a lifestyle factor, e.g typical wake time. When compared between subjects, indicative of differences in \textbf{habits}. Also referred to as `\textit{trait}' for brevity.\\\hline
         Lifestyle \textit{state} & within-subject deviation (on a given day) from their trait, indicative of \textbf{day-to-day deviations from habits} (e.g, waking up 1 hour later than usual on a given day). Also referred to as `\textit{state}' for brevity. \\\hline
    \end{tabularx}
    \caption{Definitions of terms used.}
  
\end{table}

\newpage
\section{Time-of-day variables local linearity}\label{section:supp_local_lin}

\begin{figure}[h!]
    \centering
    \begin{subfigure}[h]{1\textwidth}
        \centering
        \includegraphics[scale=1]{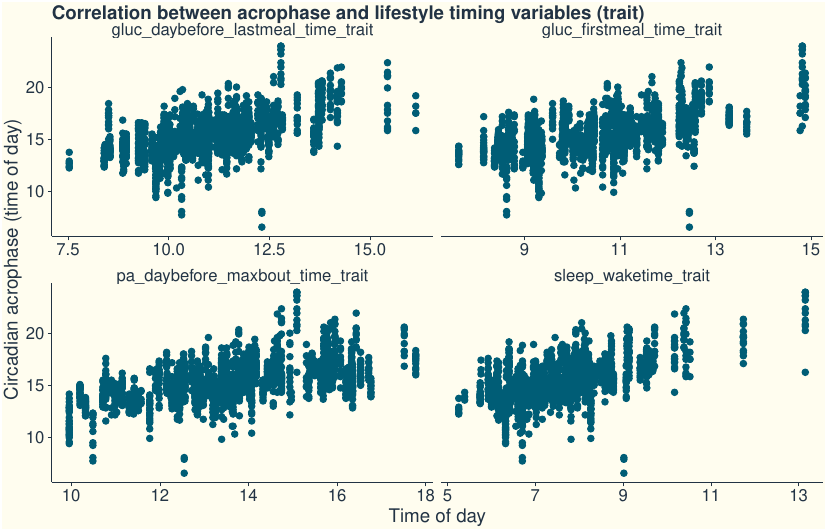}
        \caption{}\label{fig:local_lin_trait}
    \end{subfigure}
    \begin{subfigure}[h]{1\textwidth}
        \centering
        \includegraphics[scale=1]{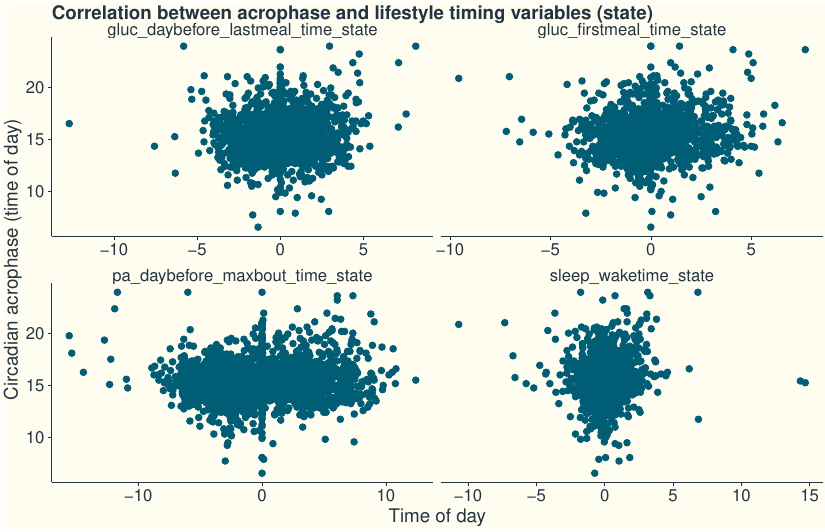}
        \caption{}\label{fig:local_lin_state}
    \end{subfigure}

    \caption{}
    
\end{figure}

A key assumption of linear mixed effects models is that predictor and response factors are linearly related. Circadian acrophase (our model response factor) as well as any timing-related trait predictors are both circular variables (with range 0.00-23.99). A circular variable is defined by `wrapping around' at minima and maxima (i.e, the value succeeding 23.99h is 0.00h, and the value preceding 0.00h is 23.99h). If one or both time-of-day factors are centered near this `midnight boundary', non-linear `wrap around' correlations are possible, which are not suitable for linear mixed effects models. However, we show in Figures \ref{fig:local_lin_trait} and \ref{fig:local_lin_state} that the relationship between timing predictors and circadian acrophase are `locally linear' (they do not cross the midnight boundary), so do not exhibit problematic wrap-around effects, and are suitable for linear modelling. Selection of predictors was motivated by this challenge: for example, we selected wake time (centered in the morning) over sleep onset time (centered near the midnight boundary). State factors are not actually circular 'time of day' variables, but reflect the deviation from a time of day (the trait) in hours. They are shown here alongside accompanying traits for reference. 

\newpage
\section{Demographics distributions}\label{section:supp_demographics}

\begin{figure}[h!]
    \centering
    \includegraphics[width=\linewidth]{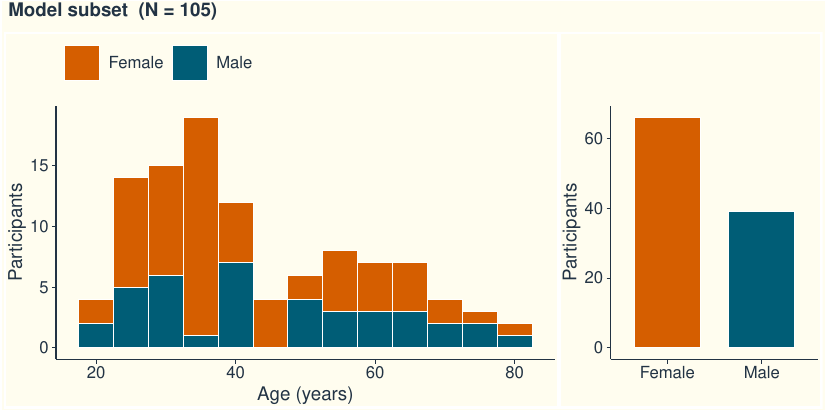}
    \caption{Age and sex distributions for the n=105 individuals included in our analysis.}
\end{figure}

\newpage
\section{Leave-one-subject-out}\label{section:supp_LOSO}

\begin{figure}[h!]
    \centering
    \includegraphics[scale=1]{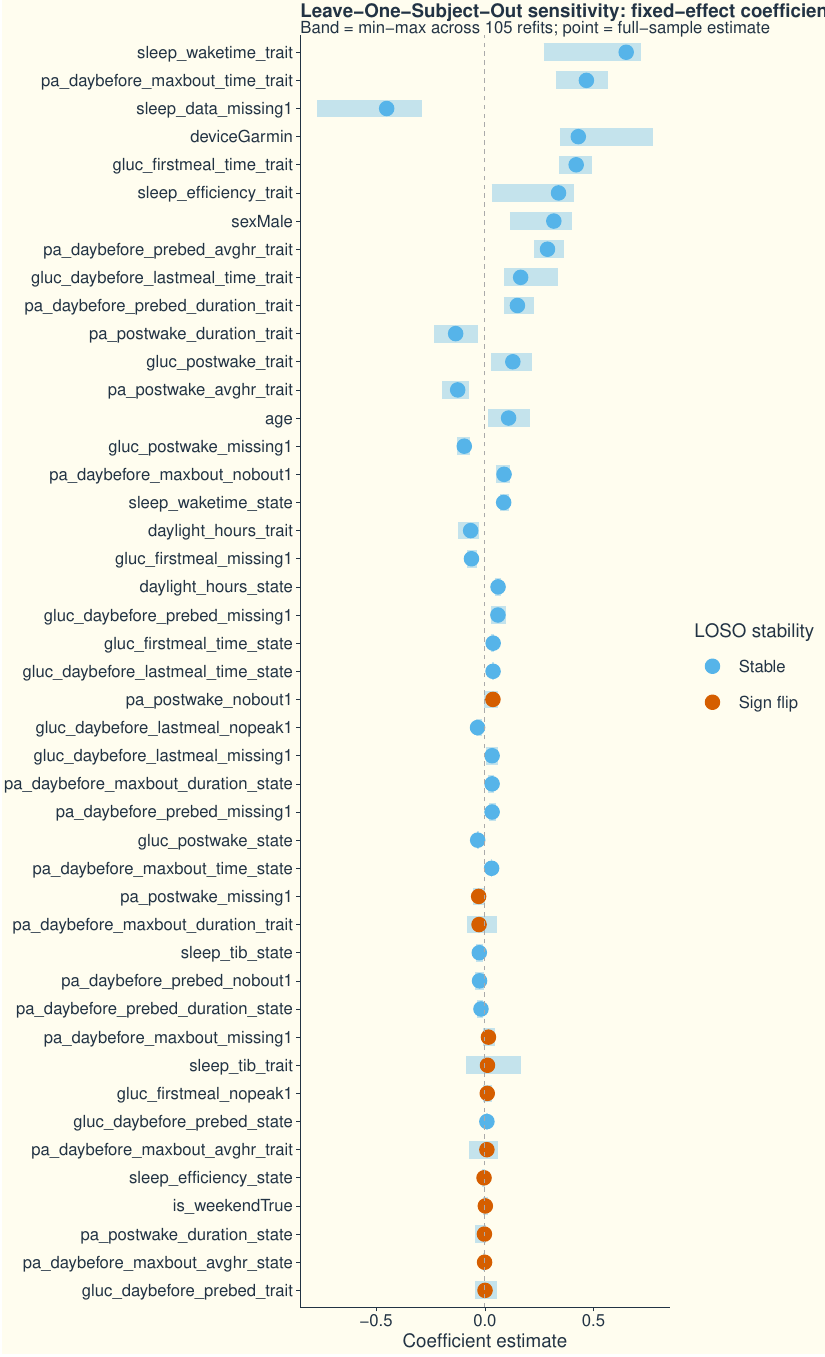}
    \caption{The distribution of coefficient estimates across 105 iterations (in each iteration, one of 105 subjects were left out, to determine confidence intervals for each factor). Points show standardised model estimates (beta weights), with 95\% confidence intervals overlaid.}\label{fig:supp_loso_sensitivity}

\end{figure}

%PURPOSE
To determine if observed results were driven by particular influential subjects, we re-ran the model fit for each subject in the sample, dropping that subject from the model fit each time. Supplementary Figure \ref{fig:supp_loso_sensitivity}, similarly to Figure \ref{fig:indi_estimates}, shows the distribution of each model coefficient estimate across all iterations. Confidence intervals (CI) reflect the consistency of that estimate across all iterations, with wider intervals reflecting that the estimate is unstable (can be driven by particular individuals). Point colour reflects whether the estimate `flipped sign' (e.g, went from positive (acrophase delay) to negative (acrophase advance)) across model fits, which would suggest that the factors' precise relationship with acrophase is unclear. 

%Findings
Factors with large estimates retained directionality.  The most influential trait factors (mean, $CI_{min}-CI_{max}$ in Figure \ref{fig:indi_estimates}  -- wake (0.65, 0.27-0.79), previous day peak-activity (0.47, 0.33-0.56), and first meal timing (0.43, 0.34-0.50) -- remained consistent in direction and reasonably stable. Sign flips occurred only for factors with smaller estimates centered on zero, so are of little concern. 

%Conclusion
Besides wake time, major traits identified (previous day peak-activity and first meal timing) remained consistent (low CI), and all retained direction. The large confidence interval for sleep timing may be reflective of inter-individual chronotype or work schedule differences. Smaller estimates remained consistently small across all iterations, providing further confidence that these factors are not influential.\newline

% Purpose
Additionally, this process facilitated estimating uncertainty for statistics reported in Figure~\ref{fig:maindonut} (semi-partial $R^2$ by category) and Figure~\ref{fig:traitstate} (semi-partial $R^2$ of trait and state factors, both overall and within categories, and snijders-bosker multi-level variance decomposition). We calculated the mean and standard deviation of these statistics across all iterations (Tables~\ref{tbl:loso_tbl_first}-\ref{tbl:loso_tbl_last}).

% Findings

% \begin{itemize}
%     \item For Figure~\ref{fig:maindonut}:
%     \begin{itemize}
%             \item sleep: 0.036 $\pm$ 0.002
%             \item food: 0.019 $\pm$ 0.001
%             \item physical activity: 0.071 $\pm$ 0.003
%             \item demograpics \& light: 0.012 $\pm$ 0.001
%             \item random intercept: 0.083 $\pm$ 0.007
%             \item unexplained: 0.393 $\pm$ 0.006
%     \end{itemize}
    
%     \item For Figure~\ref{fig:traitstate}A:
%     \begin{itemize}
%         \item trait: 0.423 $\pm$ 0.006
%         \item state: 0.0095 $\pm$ 0.0005
%     \end{itemize}
    
%     \item For Figure~\ref{fig:traitstate}B:
%     \begin{itemize}
%         \item physical activity (trait: 0.067 $\pm$ 0.003, state: 0.0012 $\pm$ 0.0001)
%         \item sleep (trait: 0.032 $\pm$ 0.002, state: 0.0028 $\pm$ 0.0002)
%         \item food (trait: 0.017 $\pm$ 0.001, state: 0.0024 $\pm$ 0.0001)
%     \end{itemize}

%     \item For Snidjers-Bosker multi-level variance decomposition: 
%     \begin{itemize}
%         \item $R^2_{between}$: 0.865 $\pm$ 0.010
%         \item $R^2_{within}$: 0.019 $\pm$ 0.007
%     \end{itemize}

% \end{itemize}

\begin{table}[h]
\centering
\caption{Figure~\ref{fig:maindonut}: Semi-partial $R^2$ by category}\label{tbl:loso_tbl_first}
\begin{tabular}{lcc}
\toprule
\textbf{Category} & \textbf{Mean} & \textbf{SD} \\
\midrule
Sleep & 0.0358 & 0.0021 \\
Food & 0.0194 & 0.0014 \\
Physical activity & 0.0705 & 0.0032 \\
Demographics \& light & 0.0118 & 0.0011 \\
Overlap & 0.3862 & 0.0090 \\
Random intercept & 0.0829 & 0.0066 \\
Unexplained & 0.3934 & 0.0059 \\
\bottomrule
\end{tabular}
\end{table}

\begin{table}[h]
\centering
\caption{Figure~\ref{fig:traitstate}A: Overall trait and state semi-partial $R^2$}
\begin{tabular}{lcc}
\toprule
\textbf{Factors} & \textbf{Mean} & \textbf{SD} \\
\midrule
Traits & 0.4234 & 0.0064 \\
States & 0.0095 & 0.0005 \\
\bottomrule
\end{tabular}
\end{table}

\begin{table}[h]
\centering
\caption{Figure~\ref{fig:traitstate}B: Trait and state semi-partial $R^2$ within categories}
\begin{tabular}{lcccc}
\toprule
\textbf{Category} & \multicolumn{2}{c}{\textbf{Traits}} & \multicolumn{2}{c}{\textbf{States}} \\
\cmidrule(lr){2-3} \cmidrule(lr){4-5}
 & \textbf{Mean} & \textbf{SD} & \textbf{Mean} & \textbf{SD} \\
\midrule
Sleep & 0.0325 & 0.0021 & 0.0028 & 0.0002 \\
Food & 0.0173 & 0.0014 & 0.0024 & 0.0001 \\
Physical activity & 0.0671 & 0.0032 & 0.0012 & 0.0001 \\
\bottomrule
\end{tabular}
\end{table}

\begin{table}[h]
\centering
\caption{Snijders-Bosker multi-level variance decomposition}\label{tbl:loso_tbl_last}
\begin{tabular}{lcc}
\toprule
\textbf{Statistic} & \textbf{Mean} & \textbf{SD} \\
\midrule
$R^2_{\text{between}}$ & 0.8655 & 0.0101 \\
$R^2_{\text{within}}$ & 0.0190 & 0.0070 \\
\bottomrule
\end{tabular}
\end{table}

% Conclusion
Point estimates were consistent with those reported in Figures~\ref{fig:maindonut}~and~\ref{fig:traitstate} on average, and stable across iterations.

\newpage
\section{Weekends vs Weekdays}\label{section:supp_weekend}

\begin{figure}[h!]
    \centering
    \includegraphics[scale=1]{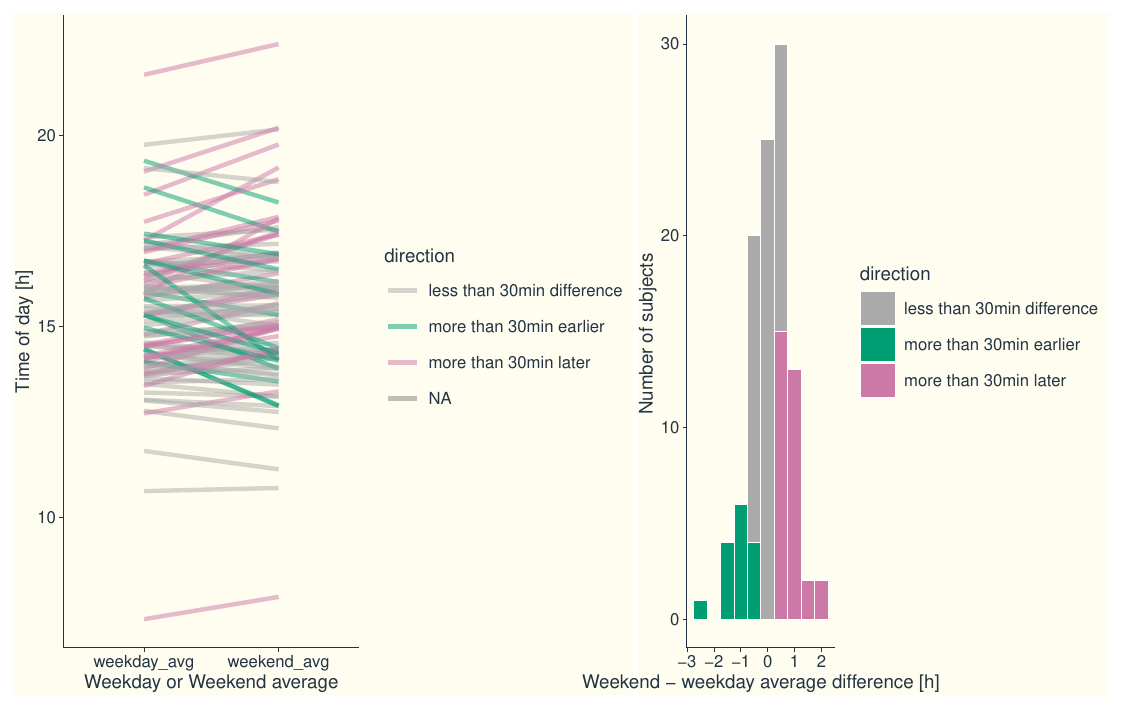}
    \caption{\textbf{A:} Difference between average circadian acrophase calculated over weekdays \textit{vs.} weekends for each subject (line), coloured by direction of difference. \textbf{B:} distribution across subjects of differences between weekday and weekend average circadian acrophase (mean=0.12h, median=0.16h).}\label{fig:supp_weekday_weekend_dist}
\end{figure}

\begin{figure}[h!]
    \centering
         
    \begin{subfigure}[h]{1\textwidth}
        \centering
        \includegraphics[scale=1]{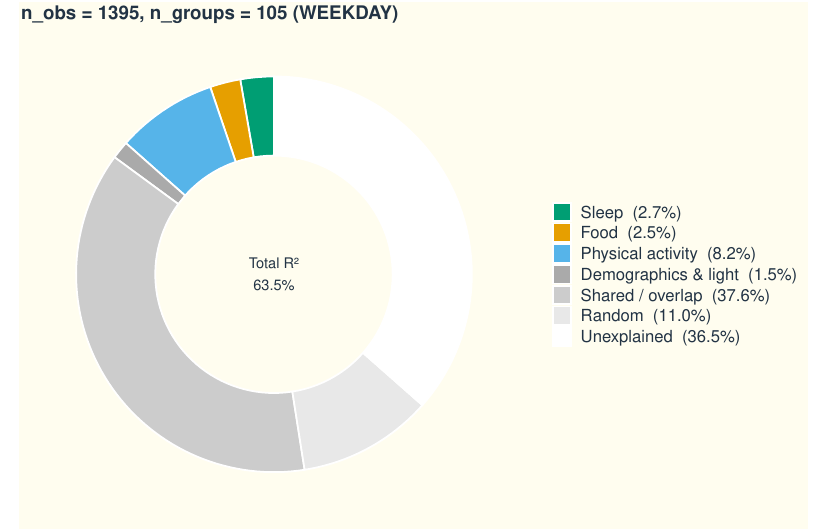}
        \caption{Model conditional and category semi-partial $R^2$s calculated using a model including only weekday cycles.}
    \end{subfigure}
    \begin{subfigure}[h]{1\textwidth}
        \centering
        \includegraphics[scale=1]{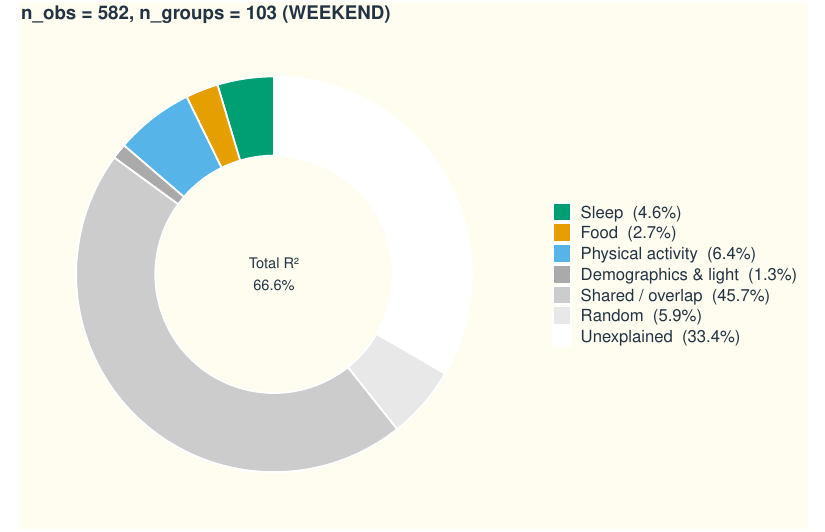}
        \caption{Model conditional and category semi-partial $R^2$s calculated using a model including only weekend cycles.}
    \end{subfigure}

    \caption{}
    \label{fig:supp_weekday_weekend_donut}
\end{figure}

\begin{figure}[h!]
    \centering
         
    \begin{subfigure}[h]{1\textwidth}
        \centering
        \includegraphics[scale=0.75]{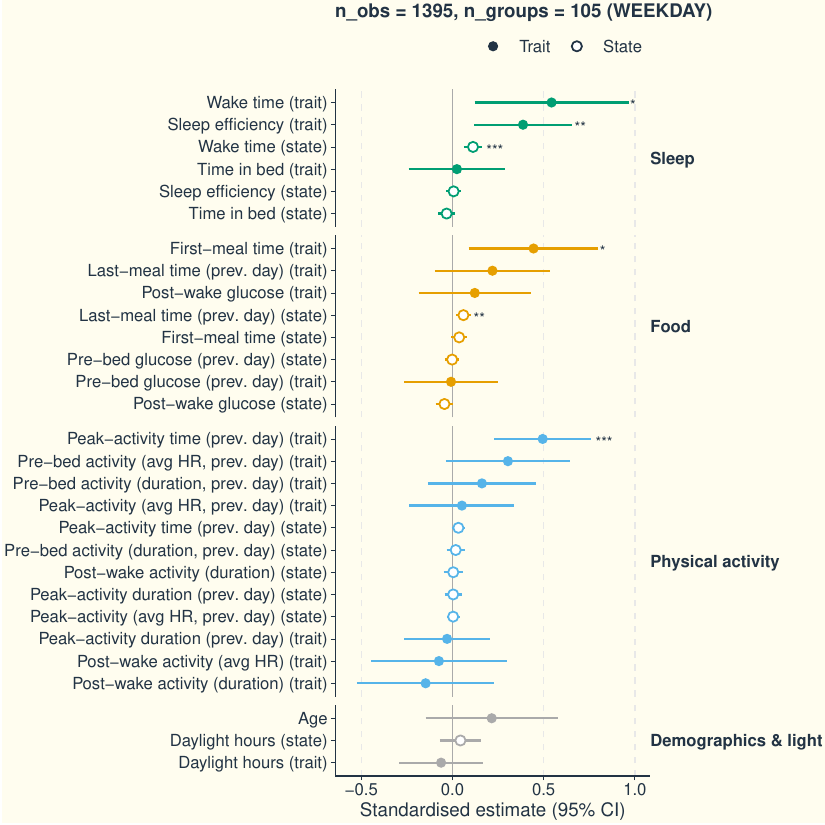}
        \caption{Model factor coefficients calculated using a model including only weekday cycles.}
    \end{subfigure}
    \begin{subfigure}[h]{1\textwidth}
        \centering
        \includegraphics[scale=0.75]{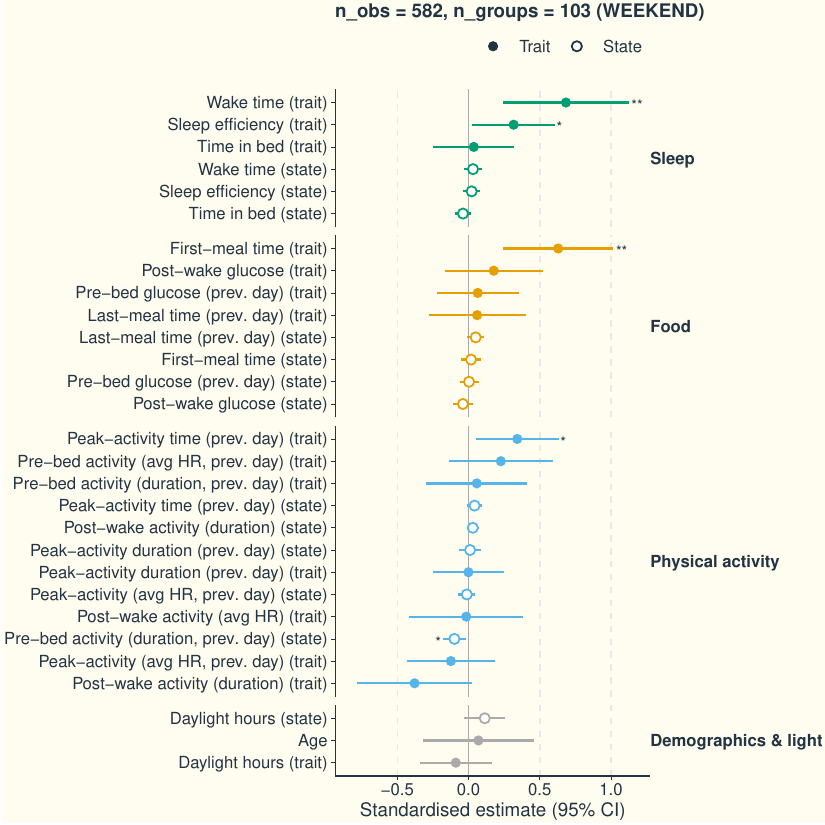}
        \caption{Model factor coefficients calculated using a model including only weekend cycles.}
    \end{subfigure}

    \caption{}
    \label{fig:supp_weekday_weekend_forest}
\end{figure}

% Finding
In our main model, there was a negligible weekday \textit{vs.} weekend effect on acrophase (`is\_weekend' factor, $\beta= 0.002 \pm 0.04$,  Supplementary Table~\ref{tab:main_coefficients}). Here, we investigate the weekend effect in more detail. Looking at weekday-weekend distributions of acrophase (Figure~\ref{fig:supp_weekday_weekend_dist}), most subjects did not have a substantial difference in acrophase between weekdays and weekends, though the number of subjects with a $\geq 30 min$ later acrophase on weekends was greater than the number with a $\geq 30 min$ earlier weekend acrophase. When re-running the model over weekdays-only and weekends-only (Figures~\ref{fig:supp_weekday_weekend_donut} and \ref{fig:supp_weekday_weekend_forest}), trends seen in Figures~\ref{fig:maindonut} and \ref{fig:indi_estimates} generally held. However, sleep and overlap explained relatively more variance during weekends, and conversely, physical activity and the random intercept explained more variance during weekdays. 
 
% Magnitude
The difference between weekend and weekday acrophase was small at 0.12 hours on average (Figure~\ref{fig:supp_weekday_weekend_dist}).
When modelling weekends and weekdays separately, during weekends sleep, overlap, physical activity and random intercept explained \textbf{4.6\%} , \textbf{45.7\%} , 6.4\%  and 5.9\% respectively (Figure~\ref{fig:supp_weekday_weekend_donut}).
Conversely, during weekdays sleep, overlap, physical activity and random intercept explained 2.7\%, 27.6\%, \textbf{8.2\%} and \textbf{11.0\%} respectively.
Other categories did not differ substantially between the two models. 

% Biological Interpretation
The trend towards later acrophase on weekends is most likely reflective of the social jetlag phenomenon. 
%
%While the difference between weekdays and weekends was minimal for most subjects, there were clear weekend-weekday differences for some subjects. Despite this, the weekend factor was not influential in the main model, implying that weekday-weekend differences can be accounted for by the modelled lifestyle factors. 
Despite clear weekend-weekday differences for some subjects, the weekend factor was not influential in the main model, implying that weekday-weekend differences can be accounted for by the modelled lifestyle factors. 
%
%When modelling weekends and weekdays separately, the greater influence of sleep on acrophase during weekends may similarly be reflective of the social jetlag phenomenon. Conversely, the greater influence of physical activity and the random intercept (reflective of between-subject differences not captured by traits) during the week may be reflective of the influence of working schedules. 

\newpage
\section{Multicollinearity assessment via variance inflation factors (VIFs)}\label{section:supp_vifs}

% Purpose
`Multicollinearity' refers to the association between model predictor factors, specifically conditional on other factors also included in the model (as opposed to a direct correlation between raw predictors). If multicollinearity between two predictors A and B is high, the value of adding B will be low (and will likely result in a low model estimate with high standard error for B), even if both A and B are strongly associated with the response factor. 

The variance inflation factor (VIF) is a measure of the multicollinearity of a predictor. While accepted thresholds vary, generally a VIF below 5 suggests low multicollinearity, between 5 and 10 suggests moderate multicollinearity, and above 10 indicates excessive multicollinearity and warrants removal of a predictor from a model \citep{ludecke_performance_2021}.

% Finding
We report VIFs for factors included in the model in Table~\ref{tbl:supp_vif}. Besides wake time (trait), most factors had low multicollinearity. 

% Conclusion
While wake time displays moderate multicollinearity, this is not high enough to warrant removal from the model. Related lifestyle factors will inherently display some multicollinearity. Post-wake activity average HR and previous day pre-bed activity average HR \textbf{states} (not shown here) were previously dropped for excessive VIFs so were not included in the model, but their trait factors were kept.

 \begin{table}[h!]
        \centering

        \scalebox{0.70}{
            \begin{tabular}{l|r|r|r|r|r|r|r}
                \hline
                Factor & VIF & VIF\_CI\_low & VIF\_CI\_high & SE\_factor & Tolerance & Tolerance\_CI\_low & Tolerance\_CI\_high\\
                \hline
                sleep\_waketime\_trait & 6.24 & 5.77 & 6.76 & 2.50 & 0.16 & 0.15 & 0.17\\
                \hline
                pa\_postwake\_duration\_trait & 4.66 & 4.32 & 5.04 & 2.16 & 0.21 & 0.20 & 0.23\\
                \hline
                gluc\_firstmeal\_time\_trait & 4.41 & 4.09 & 4.77 & 2.10 & 0.23 & 0.21 & 0.24\\
                \hline
                pa\_postwake\_avghr\_trait & 4.38 & 4.06 & 4.73 & 2.09 & 0.23 & 0.21 & 0.25\\
                \hline
                age & 4.04 & 3.75 & 4.36 & 2.01 & 0.25 & 0.23 & 0.27\\
                \hline
                pa\_daybefore\_prebed\_avghr\_trait & 3.80 & 3.53 & 4.10 & 1.95 & 0.26 & 0.24 & 0.28\\
                \hline
                pa\_daybefore\_prebed\_duration\_trait & 3.63 & 3.38 & 3.92 & 1.91 & 0.28 & 0.26 & 0.30\\
                \hline
                gluc\_daybefore\_lastmeal\_time\_trait & 3.59 & 3.34 & 3.87 & 1.89 & 0.28 & 0.26 & 0.30\\
                \hline
                gluc\_postwake\_trait & 3.32 & 3.09 & 3.58 & 1.82 & 0.30 & 0.28 & 0.32\\
                \hline
                device & 2.87 & 2.67 & 3.08 & 1.69 & 0.35 & 0.32 & 0.37\\
                \hline
                pa\_daybefore\_maxbout\_avghr\_trait & 2.65 & 2.48 & 2.85 & 1.63 & 0.38 & 0.35 & 0.40\\
                \hline
                sleep\_efficiency\_trait & 2.61 & 2.43 & 2.80 & 1.61 & 0.38 & 0.36 & 0.41\\
                \hline
                pa\_daybefore\_maxbout\_time\_trait & 2.27 & 2.12 & 2.43 & 1.51 & 0.44 & 0.41 & 0.47\\
                \hline
                gluc\_daybefore\_lastmeal\_missing & 2.27 & 2.12 & 2.43 & 1.51 & 0.44 & 0.41 & 0.47\\
                \hline
                gluc\_daybefore\_prebed\_trait & 2.19 & 2.05 & 2.34 & 1.48 & 0.46 & 0.43 & 0.49\\
                \hline
                gluc\_daybefore\_prebed\_missing & 2.14 & 2.01 & 2.29 & 1.46 & 0.47 & 0.44 & 0.50\\
                \hline
                sleep\_tib\_trait & 2.11 & 1.98 & 2.25 & 1.45 & 0.47 & 0.44 & 0.51\\
                \hline
                pa\_daybefore\_maxbout\_duration\_trait & 1.67 & 1.57 & 1.78 & 1.29 & 0.60 & 0.56 & 0.64\\
                \hline
                pa\_daybefore\_prebed\_duration\_state & 1.63 & 1.54 & 1.74 & 1.28 & 0.61 & 0.58 & 0.65\\
                \hline
                daylight\_hours\_trait & 1.61 & 1.52 & 1.71 & 1.27 & 0.62 & 0.58 & 0.66\\
                \hline
                sleep\_waketime\_state & 1.52 & 1.44 & 1.61 & 1.23 & 0.66 & 0.62 & 0.70\\
                \hline
                pa\_daybefore\_prebed\_nobout & 1.51 & 1.43 & 1.60 & 1.23 & 0.66 & 0.62 & 0.70\\
                \hline
                pa\_daybefore\_maxbout\_duration\_state & 1.42 & 1.35 & 1.50 & 1.19 & 0.71 & 0.66 & 0.74\\
                \hline
                gluc\_postwake\_missing & 1.41 & 1.34 & 1.50 & 1.19 & 0.71 & 0.67 & 0.75\\
                \hline
                pa\_postwake\_duration\_state & 1.39 & 1.32 & 1.47 & 1.18 & 0.72 & 0.68 & 0.76\\
                \hline
                sex & 1.37 & 1.30 & 1.46 & 1.17 & 0.73 & 0.69 & 0.77\\
                \hline
                sleep\_tib\_state & 1.33 & 1.27 & 1.41 & 1.15 & 0.75 & 0.71 & 0.79\\
                \hline
                gluc\_firstmeal\_time\_state & 1.32 & 1.26 & 1.40 & 1.15 & 0.76 & 0.71 & 0.79\\
                \hline
                gluc\_firstmeal\_missing & 1.32 & 1.25 & 1.40 & 1.15 & 0.76 & 0.72 & 0.80\\
                \hline
                pa\_daybefore\_prebed\_missing & 1.28 & 1.22 & 1.35 & 1.13 & 0.78 & 0.74 & 0.82\\
                \hline
                pa\_postwake\_nobout & 1.27 & 1.21 & 1.35 & 1.13 & 0.79 & 0.74 & 0.82\\
                \hline
                gluc\_postwake\_state & 1.12 & 1.08 & 1.19 & 1.06 & 0.89 & 0.84 & 0.93\\
                \hline
                pa\_postwake\_missing & 1.11 & 1.07 & 1.17 & 1.05 & 0.90 & 0.85 & 0.94\\
                \hline
                gluc\_daybefore\_lastmeal\_time\_state & 1.11 & 1.06 & 1.17 & 1.05 & 0.90 & 0.85 & 0.94\\
                \hline
                is\_weekend & 1.10 & 1.06 & 1.17 & 1.05 & 0.91 & 0.85 & 0.94\\
                \hline
                gluc\_daybefore\_prebed\_state & 1.10 & 1.06 & 1.17 & 1.05 & 0.91 & 0.86 & 0.94\\
                \hline
                gluc\_daybefore\_lastmeal\_nopeak & 1.08 & 1.04 & 1.15 & 1.04 & 0.93 & 0.87 & 0.96\\
                \hline
                sleep\_efficiency\_state & 1.08 & 1.04 & 1.15 & 1.04 & 0.93 & 0.87 & 0.96\\
                \hline
                pa\_daybefore\_maxbout\_time\_state & 1.06 & 1.03 & 1.13 & 1.03 & 0.94 & 0.88 & 0.97\\
                \hline
                sleep\_data\_missing & 1.04 & 1.01 & 1.13 & 1.02 & 0.96 & 0.89 & 0.99\\
                \hline
                pa\_daybefore\_maxbout\_missing & 1.04 & 1.01 & 1.13 & 1.02 & 0.96 & 0.89 & 0.99\\
                \hline
                gluc\_firstmeal\_nopeak & 1.03 & 1.01 & 1.13 & 1.02 & 0.97 & 0.88 & 0.99\\
                \hline
                pa\_daybefore\_maxbout\_avghr\_state & 1.03 & 1.01 & 1.14 & 1.02 & 0.97 & 0.88 & 0.99\\
                \hline
                pa\_daybefore\_maxbout\_nobout & 1.03 & 1.00 & 1.15 & 1.01 & 0.97 & 0.87 & 1.00\\
                \hline
                daylight\_hours\_state & 1.02 & 1.00 & 1.28 & 1.01 & 0.99 & 0.78 & 1.00\\
                \hline
            \end{tabular}
 
    }
    \caption{Ordered VIFs for the main model factors, derived using the R function performance::check\_collinearity() \citep{ludecke_performance_2021}\label{tbl:supp_vif}}
    
\end{table}

\newpage
\section{Interaction terms}\label{section:supp_interactions}

% Purpose
Interaction terms were not included in the main model due to sample size restrictions. Here, we tested models with additional interaction terms between each pair of predictors. With n=45 predictors (Table~\ref{tab:main_coefficients}), 990 ($n(n-1)/2$) models were tested. 
%  Finding
The top 20 (ordered by increase in marginal $R^2$) are shown in Table~\ref{tab:interactions}. The highest addition to model marginal $R^2$ was 0.03 (3\%) by adding an interaction term between sleep efficiency and previous day pre-bed activity duration traits. 
% Conclusion
Adding pair-wise interaction terms individually did not substantially improve model fit. This may suggest that interactions between lifestyle factors under natural circumstances do not play a role in modulating CRHR acrophase, though this requires further investigation. Interestingly, interactions between traits and states did not emerge: implying that, for example, the influence of the change from typical wake time on acrophase is not different for early or late wakers.

 \begin{table}[h!]
            \centering

            \scalebox{0.70}{
                
                \begin{tabular}{l|l|r|r|r|r|r|r}
                    \hline
                    var1 & var2 & R2m\_base & R2c\_base & R2m\_new & R2c\_new & delta\_R2m & delta\_R2c\\
                    \hline
                    sleep\_efficiency\_trait & pa\_daybefore\_prebed\_duration\_trait & 0.523 & 0.607 & 0.553 & 0.616 & 0.030 & 0.010\\
                    \hline
                    sleep\_waketime\_trait & gluc\_daybefore\_prebed\_trait & 0.523 & 0.607 & 0.542 & 0.605 & 0.018 & -0.001\\
                    \hline
                    age & sex & 0.523 & 0.607 & 0.537 & 0.610 & 0.014 & 0.004\\
                    \hline
                    sleep\_waketime\_trait & gluc\_postwake\_trait & 0.523 & 0.607 & 0.536 & 0.604 & 0.013 & -0.003\\
                    \hline
                    device & pa\_daybefore\_prebed\_duration\_trait & 0.523 & 0.607 & 0.536 & 0.609 & 0.012 & 0.002\\
                    \hline
                    daylight\_hours\_trait & gluc\_postwake\_trait & 0.523 & 0.607 & 0.534 & 0.604 & 0.011 & -0.003\\
                    \hline
                    gluc\_firstmeal\_time\_trait & pa\_daybefore\_maxbout\_avghr\_trait & 0.523 & 0.607 & 0.533 & 0.608 & 0.010 & 0.002\\
                    \hline
                    gluc\_postwake\_trait & gluc\_firstmeal\_time\_trait & 0.523 & 0.607 & 0.533 & 0.600 & 0.010 & -0.007\\
                    \hline
                    gluc\_firstmeal\_time\_trait & pa\_daybefore\_prebed\_duration\_trait & 0.523 & 0.607 & 0.533 & 0.609 & 0.009 & 0.003\\
                    \hline
                    sleep\_waketime\_trait & pa\_daybefore\_maxbout\_avghr\_trait & 0.523 & 0.607 & 0.532 & 0.609 & 0.008 & 0.002\\
                    \hline
                    sleep\_waketime\_trait & pa\_daybefore\_maxbout\_missing & 0.523 & 0.607 & 0.532 & 0.610 & 0.008 & 0.004\\
                    \hline
                    gluc\_daybefore\_prebed\_trait & gluc\_daybefore\_lastmeal\_time\_trait & 0.523 & 0.607 & 0.531 & 0.605 & 0.008 & -0.002\\
                    \hline
                    gluc\_daybefore\_prebed\_trait & pa\_daybefore\_maxbout\_avghr\_trait & 0.523 & 0.607 & 0.531 & 0.602 & 0.008 & -0.005\\
                    \hline
                    sleep\_waketime\_trait & sleep\_efficiency\_trait & 0.523 & 0.607 & 0.531 & 0.611 & 0.008 & 0.005\\
                    \hline
                    gluc\_postwake\_trait & pa\_postwake\_avghr\_trait & 0.523 & 0.607 & 0.531 & 0.607 & 0.007 & 0.000\\
                    \hline
                    gluc\_daybefore\_prebed\_trait & pa\_postwake\_avghr\_trait & 0.523 & 0.607 & 0.531 & 0.605 & 0.007 & -0.001\\
                    \hline
                    sleep\_efficiency\_trait & gluc\_firstmeal\_time\_trait & 0.523 & 0.607 & 0.530 & 0.611 & 0.007 & 0.005\\
                    \hline
                    gluc\_daybefore\_prebed\_trait & gluc\_firstmeal\_time\_trait & 0.523 & 0.607 & 0.530 & 0.602 & 0.007 & -0.004\\
                    \hline
                    sex & pa\_daybefore\_maxbout\_time\_trait & 0.523 & 0.607 & 0.530 & 0.610 & 0.006 & 0.004\\
                    \hline
                    sleep\_efficiency\_trait & gluc\_postwake\_trait & 0.523 & 0.607 & 0.530 & 0.604 & 0.006 & -0.002\\
                    \hline
                \end{tabular}
                                
        }
        \caption{The change in marginal ($mR^2$) and conditional ($cR^2$) when each pair of predictor factors in the model were added as an interaction term (var1:var2). `\_base' refers to the marginal and conditional $R^2$ of the original model. `\_new' refers to the marginal and conditional $R^2$ of a model with this interaction term added. `delta' refers to either the marginal or conditional $R^2$ gain by adding this interaction term (e.g $R^2$m\_new - $R^2$m\_base). Results are ordered by marginal $R^2$ gained (delta\_$R^2$m). Top 20 models shown.}\label{tab:interactions}
        
        \end{table}

\newpage
% \bibliographystylesupp{plainnat}
% \bibliographysupp{supplementary}
\printbibliography[title={Supplementary References}]
\end{refsection}

\end{document}